\documentclass[fleqn,usenatbib]{mnras}
\usepackage{newtxtext,newtxmath}
\usepackage[T1]{fontenc}
\DeclareRobustCommand{\VAN}[3]{#2}
\let\VANthebibliography\thebibliography
\def\thebibliography{\DeclareRobustCommand{\VAN}[3]{##3}\VANthebibliography}
\usepackage{subcaption}
\usepackage{multicol}
\usepackage{graphicx}
\usepackage{amsmath}
\usepackage{xcolor}
\usepackage{float}
\usepackage{afterpage}
\usepackage{caption}
\usepackage{multirow}

\title[Time-Delay Recovery via Mutual Information]{Time-Delay Recovery in AGN Reverberation Mapping via Mutual Information}

\author[R. A. A. Souza et al.]{
R. A. A. Souza$^{1,2}$, E. Hatziminaoglou$^{3,4,5}$, D. O. Fontinele$ ^{6}$, L. B. T. Santos$^{7,8}$, M.V.D. Souza$^{9}$, T. Stosic$^{10}$\\
$^{1}$Physics Program, Universidade de Pernambuco, Campus Mata Norte, Nazar\'{e} da Mata, PE 55800-000, Brazil \\
$^{2}$Departamento de F\'{i}sica, Universidade Federal Rural de Pernambuco, Recife, PE, 50670-901, Brazil \\
$^{3}$Instituto de Astrof\'{i}sica de Canarias, 38205 La Laguna, Tenerife, Spain \\
$^{4}$Departamento de Astrof\'{i}sica, Universidad de La Laguna, 38206 La Laguna, Tenerife, Spain\\
$^{5}$ESO, Karl-Schwarzschild-Str. 2, 85748 Garching bei M\"unchen, Germany \\
$^{6}$ Departamento de F\'isica Te\'orica e Experimental, Universidade Federal do Rio Grande do Norte, Campus Universit\'ario, Natal, RN, 59072-970, Brazil\\
$^{7}$Group of Orbital \& Rotational Research in Irregular Objects \& Observational astroNomy (ORION), UNESP, Guaratinguet\'{a}, 12516-410, S\~{a}o Paulo, Brazil\\
$^{8}$PostGrad Program in Systems Engineering (PPGES), Universidade de Pernambuco (UPE), Recife, 50720-001, Brazil\\
$^{9}$Departamento de Química Fundamental, Universidade Federal de Pernambuco, 50670-901 Recife-PE, Brazil\\
$^{10}$Departamento de Estat\'{i}stica e Inform\'{a}tica, Universidade Federal Rural de Pernambuco, Recife, PE, 50670-901, Brazil\\
}

\date{Accepted . Received 2; in original form .}
\pubyear{2026}

\begin{document}
\label{firstpage}
\pagerange{\pageref{firstpage},\pageref{lastpage}}
\maketitle

\begin{abstract}
Active galactic nuclei exhibit complex, nonlinear multiwavelength
variability that standard cross-correlation techniques, sensitive only
to linear covariance, may fail to capture fully.  We apply time-lagged
Mutual Information (MI) lag estimation, via the
Kraskov-Stoegbauer-Grassberger $k$-nearest-neighbour estimator
($k=5$) with a 150-iteration Random Subset Selection bootstrap, to the
light curves of four Seyfert galaxies (Mrk~509, NGC~5548, NGC~4151, and
NGC~4593) monitored across eight energy bands by \textit{Swift}. 
%as part of the \textit{Swift} intensive disc reverberation mapping survey of \citet{edelson2019first}. 
In the UV/optical regime the MI estimator
successfully reproduces the inter-band disc reprocessing lags
established by the interpolated cross-correlation function, with
performance degrading toward the optical where low variability
amplitude produces poorly defined MI peaks.  In the X-ray bands,
where the physical X-ray, UV coupling is known to be weak
%\citep{edelson2019first}, 
the estimator fails to recover consistent
lags across the sample, with the notable exception of NGC~4593.
Contextualising these outcomes within the multifractal
characterisation of our previous work,
%\citet{Souza2026RM}, 
we discuss how MI performance
relates to the multifractal spectral width $\Delta\alpha$ of each
band, and we propose $\Delta\alpha$ as a candidate diagnostic for anticipating where
MI-based reverberation mapping is likely to succeed or fail in 
multiband datasets of active galactic nuclei.
\end{abstract}


% ==================================================================
\section{Introduction}
\label{sec:intro}
% ==================================================================

Active galactic nuclei (AGNs) are among the most energetically extreme
objects in the Universe, powered by accretion of matter onto supermassive
black holes at the centres of their host galaxies
\citep[e.g.][]{peterson1997introduction, netzer2013physics}.  Their
emission spans the entire electromagnetic spectrum, from radio waves to
hard X-rays \citep{ulrich1997variability, shang2011next}, and varies on
timescales ranging from hours to years \citep{vaughan2003characterizing}.
This variability encodes fundamental information about the geometry,
kinematics, and physical conditions of the emitting regions surrounding
the central engine \citep{ulrich1997variability}.  Understanding these
regions, in particular the accretion disc and the broad-line region
(BLR), is crucial for constraining models of black hole growth and AGN
feedback across cosmic time \citep{kormendy2013coevolution,
heckman2014coevolution}.  

Reverberation mapping \citep{Blandford1982, Peterson1993} is one of the
most powerful observational techniques for probing the inner structure
of AGNs.  Variations in the ionising continuum emission propagate outward and are
reprocessed by surrounding gas and dust, producing delayed echoed signals arising from the finite travel time of light across the system,
at longer wavelengths.  By measuring the time lag, $\tau$, between a
reference and a subsidiary band, one can infer the light-travel distance
to the reprocessing region, mapping the spatial structure of the AGN on
scales inaccessible to direct imaging \citep{cackett2021reverberation}.
In the scenario of thermal reprocessing of extreme UV/X-ray photons to optical/UV photons by the accretion disc %disc-reprocessing framework 
\citep{cackett2007testing}, the
observed lags follow the characteristic wavelength-dependent scaling
$\tau \propto \lambda^{4/3}$, a direct prediction of the standard
thin-disc model \citep{shakura1973black}.  This scenario is supported by
X-ray reverberation lags observed in the Fourier domain of individual
Seyfert galaxies \citep{uttley2014x, kara2016global, fabian2009broad},
and by dedicated multiwavelength monitoring campaigns spanning X-ray,
UV, and optical bands simultaneously \citep{edelson2015space,
edelson2017swift, edelson2019first, mchardy2018x}.  These campaigns
have provided tight constraints on the accretion disc structure and the
geometry of the corona, while also revealing discrepancies between
observed lag amplitudes and standard thin-disc predictions, pointing towards a contribution from the BLR and diffuse continuum emission to the observed lag amplitudes \citep{cackett2018accretion}.

The standard approach to measuring inter-band lags relies on the
interpolated cross-correlation function \citep[ICCF;][]{Gaskell1987,
Peterson1998} or its discrete counterpart \citep[DCF;][]{Edelson1988}.
More sophisticated methods, such as \textsc{javelin} \citep{Zu2011},
model the light curves as damped random walk processes and infer lags
through maximum-likelihood fitting.  While these approaches have yielded
significant results, they share a common limitation: they are
fundamentally based on second-order statistics, i.e., the linear
covariance between bands.  AGN Light Curves, however, are inherently
non-Gaussian and nonlinear \citep{Vio1992, Longo1996, Provenzale1994,
Green1999}, and exhibit complex temporal behaviour that includes
intermittency \citep{Chian1997} and non-linear variability signatures
incompatible with simple stochastic models
\citep{vaughan2003characterizing}.

In particular, their variability has been shown to exhibit multifractal
scaling in lensed quasars \citep{belete2018multifractality,
belete2019novel, belete2019cosmological, assis2024multifractality,
Souza2026_MRK421} and in multiband Seyfert emission \citep{Souza2026RM}. The central measure in multifractal analysis is the width of the H\"older exponent distribution, $\Delta\alpha$, a positive quantity that is related to monofractality when $\Delta\alpha \to 0$, or to the presence of multiple fractal structures, i.e., multifractality, when $\Delta\alpha$ deviates from zero. Each H\"older exponent $\alpha(q)$ is derived from the local scaling relation between the logarithm of the fluctuation function $F_q(s)$, analogous to a generalised variance, and the logarithm of the corresponding time scale $s$. By varying the order $q$, which controls the relative weight given to large and small fluctuations, one obtains a family of scaling exponents that, through a Legendre transform, map onto the corresponding $\alpha(q)$ values composing the multifractal spectrum. When this scaling exponent is approximately independent of $q$, the spectrum collapses to a narrow range and $\Delta\alpha$ tends to, or equals, zero, indicating monofractal behaviour; when the scaling exponent instead varies with $q$, multiple H\"older exponents emerge, characterising a multifractal behaviour \citep{kantelhardt2002multifractal,Souza2026RM,SouzaMRK421}. For $q>0$, $F_q(s)$ is dominated by large fluctuations, whereas for $q<0$ it is dominated by small fluctuations. A left-skewed spectrum indicates dominance of small fluctuations, a right-skewed spectrum indicates dominance of large fluctuations, and a symmetric spectrum indicates comparable contributions from both. This asymmetry is quantified by the parameter $A$, with $A>1$, $A<1$, and $A=1$ corresponding respectively to left-skewed, right-skewed, and symmetric spectra \citep{ihlen2012introduction}.

In parallel, information theory, originally formulated by \citet{shannon1948} in the
context of communication systems, provides a natural framework for
characterising statistical dependence in a model-free, nonlinear
fashion. Unlike second-order statistics, information-theoretic measures
capture any form of dependence between two variables, linear or
nonlinear, without assuming a specific functional form or a particular
noise model \citep{cover2006}. Practical estimation from finite,
irregularly sampled data is made tractable by nonparametric density
estimators that require no binning or bandwidth choice. Using such a
measure as a lag estimator, by maximising it as a function of a trial
time shift $\tau$, yields a natural generalisation of the Cross-Correlation Function (CCF) that
imposes no linearity assumption.

In the context of astronomical time-delay
estimation, \citet{Chakraborty2022} introduced a Discrete Mutual
Information Function (DMIF) as a nonlinear alternative to the
discrete correlation function, and used toy models of accreting
sources, including X-ray binaries and AGNs, to demonstrate that
the DMIF recovers time lags in regimes where the linear DCF is biased or fails outright. The present work can be regarded as a direct extension of that proof-of-concept to real, multi-band AGNs light curves: rather than synthetic data, we apply a Kraskov-Stoegbauer-Grassberger (KSG) $k$-nearest-neighbour MI estimator to the \textit{Swift} \textit{Infrared Detected Reverberation Mapping} (IDRM) light curves of four Seyfert AGNs, benchmarking  the recovered lags against the established ICCF and \textsc{javelin} estimates of \citet{edelson2019first}, providing an observational test of the DMIF approach advocated by \citet{Chakraborty2022}.  \citet{Chelouche2017} explored randomness measures as a further alternative reverberation mapping estimator.
More recently, Gaussian-process-based cross-correlation
\citep{PozoNunez2023} and hierarchical Bayesian frameworks
\citep{Brewer2014} have further expanded the methodological landscape. Despite its broad applicability, MI-based lag estimation in AGN
reverberation mapping remains largely unexplored.

  The present work addresses this gap
directly.  We apply time-lagged MI estimation, using the
Kraskov-Stoegbauer-Grassberger \citep[KSG;][]{kraskov2004estimating}
$k$-nearest-neighbour estimator, as implemented in the
\textsc{scikit-learn} library, to the
multiwavelength light curves of four well-studied Seyfert galaxies
monitored simultaneously across eight energy bands as part of the
first \textit{Swift} intensive AGN accretion disc reverberation
mapping survey \citep{edelson2019first}, which extended earlier
pilot campaigns on NGC~5548 \citep{edelson2015space} and NGC~4151
\citep{edelson2017swift}.  Black hole masses are drawn from the AGN
Black Hole Mass Database \citep{2015BentzBlackHole}.  The estimated
lags are compared against the ICCF and \textsc{javelin} values of
\citet{edelson2019first}. 

Our analysis is further contextualised
within the multifractal and complexity characterisation of the same
dataset performed by \citealt{Souza2026RM}, who applied
Multifractal Detrended Moving Average analysis \citep{gu2010detrending} and the Fisher-Shannon
information plane \citep{vignat2003} to the identical
light curves, finding that all bands exhibit persistent long-range correlations, meaning that an increase (decrease) in the signal is more likely to be followed by another increase (decrease), whereas an antipersistent signal would tend to alternate between increases and decreases (Hurst exponent $H_{\rm Hurst} > 0.5$, where $H_{\rm Hurst} = 0.5$
corresponds to uncorrelated noise, $H_{\rm Hurst} > 0.5$ to persistence, and $H_{\rm Hurst} < 0.5$ to anti-persistence), alongside a superlinear scaling between multifractal width and fractional variability across
all four sources. 

This paper is organised as follows.  Section~\ref{sec:observationaldata}
describes the observational dataset and the fundamental properties of
the four AGNs.  Section~\ref{sec:method} presents the methodology,
covering the time-lagged MI estimator, the KSG $k$-nearest-neighbour
scheme, the light-curve pre-processing pipeline, and the Random Subset
Selection bootstrap \citep{efron1979bootstrap}.  The results and their comparison to the results of other methods are presented in Section~\ref{sec:results}, including the
MI lag curves for all sources, a comparison with the ICCF and
\textsc{javelin} values of \citet{edelson2019first}, and a discussion
contextualised within the multifractal signatures of
\citet{Souza2026RM}.  Section~\ref{sec:conclusion} summarises our main
conclusions.

% ==================================================================
\section{The AGN sample}
\label{sec:observationaldata}
% ==================================================================
 
\begin{figure*}
    \centering
    \includegraphics[width=13cm]{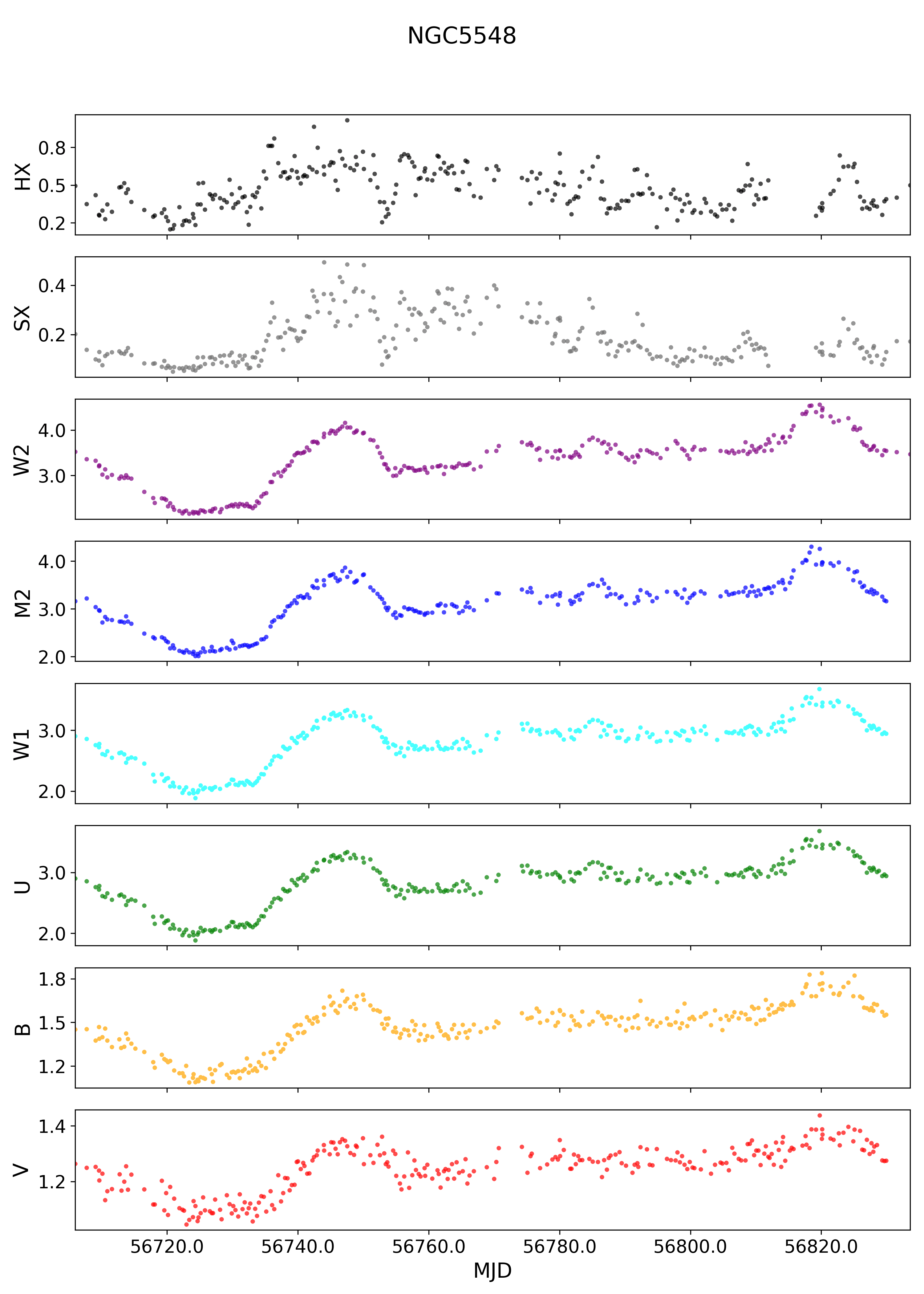}
    \caption{Light curves of NGC~5548 across all eight observational
    bands, ordered by wavelength from top to bottom. The top two panels
    show XRT observations in the HX (1.5,10~keV) and SX (0.3,1.5~keV)
    bands, while the lower six panels correspond to the \textit{Swift}/UVOT 
filters W2 ($\lambda_{\rm eff} = 1928$~\AA), M2 ($\lambda_{\rm eff} 
= 2246$~\AA), W1 ($\lambda_{\rm eff} = 2600$~\AA), U ($\lambda_{\rm 
eff} = 3465$~\AA), B ($\lambda_{\rm eff} = 4392$~\AA), and V 
($\lambda_{\rm eff} = 5468$~\AA). Each panel displays flux as a 
function of MJD, with a shared time axis and independent flux scales. X-ray
    fluxes are given in counts~s$^{-1}$, and UVOT fluxes in units of
    $10^{-14}$\,erg\,cm$^{-2}$\,s$^{-1}$\,\AA$^{-1}$.}
    \label{fig:lc_plot}
\end{figure*}

This study investigates the variability properties of four
well-studied AGNs, Mrk~509, NGC~5548, NGC~4151, and NGC~4593,
each observed by \textit{Swift} in eight energy bands within a single monitoring epoch as part of the intensive disc reverberation mapping survey 
\citet{edelson2019first}. An
example of these light curves is shown for NGC~5548 in
Fig.~\ref{fig:lc_plot}. 
%The sample selection followed three criteria: the light curves come from a publicly available, uniformly reduced dataset \citep{edelson2019first}; each light curve contains more than 150 data points, satisfying the minimum requirement for the methodsused here; and all sources were monitored in the same eight bands, enabling consistent cross-source comparison.
The sample consists of light curves with a minimum of 150 data points, meeting the minimum requirement for the method used here. It is restricted to sources observed in the same eight photometric bands to ensure consistent cross-source comparison.

The multiwavelength variability of these AGNs reflects distinct
physical mechanisms. In the Seyfert~1 galaxy Mrk~509
\citep{phillips1983outflow}, a direct correlation between soft X-ray
and UV flux variations has been reported, interpreted as variability intrinsic to the AGN, driven by changes in the ionising continuum illuminating the surrounding gas \citep{kaastra2011}.
A companion multifractal analysis of the same light curves
\citep{Souza2026RM} finds that all bands of Mrk~509 except B and V
are multifractal, with the complexity most pronounced in the X-ray bands (($\Delta\alpha = 0.321$ for \textit{Soft X-ray} (SX) and $0.216$ for \textit{Hard X-ray} (HX)), suggesting
that the innermost regions of the accretion flow drive the most
complex variability in this source.
NGC~5548, among the most thoroughly studied Seyfert~1 galaxies since
its original classification by \citet{seyfert1979nuclear}, has been
monitored since the 1960s \citep{mehdipour2015}; rapid X-ray spectral
variations have been linked to changes in the ionisation state of
intervening absorbers, while infrared coronal line variability points
to an X-ray-heated wind as a relevant energy-transport mechanism
\citep{mehdipour2015, landt2015}. \citet{Souza2026RM} report that
NGC~5548 shows the broadest multifractal spectrum in the full sample
in SX ($\Delta\alpha = 0.681$), the only band with a near-antipersistent
Hurst exponent ($H = 0.015$), alongside strong right-skewed
multifractality across its UV bands ($A \gtrsim 2$).
 
For the Seyfert~1 galaxy NGC~4151 \citep{zoghbi2019revisiting},
long-baseline observations have revealed asymmetric Balmer line
profiles, indicating a BLR with multiple kinematically distinct
components \citep{shapovalova2010}, while the historical B-band light
curve shows nonlinear, intermittent behaviour consistent with complex
underlying variability mechanisms \citep{Longo1996}. Notably, \citet{Souza2026RM} finds that NGC~4151 is the only source 
in the sample for which a single band, HX, exhibits clear multifractal  behaviour, with $\Delta\alpha = 0.542$ and a left-skewed spectrum  ($A = 0.264$), indicating that the variability is dominated by 
large-amplitude fluctuations corresponding to stronger singularities 
with smaller H\"{o}lder exponents, while all other bands are consistent with near-monofractal behaviour ($\Delta\alpha \lesssim 0.04$). The Seyfert~1
galaxy NGC~4593 \citep{cackett2018accretion} showed UV variability over
a ten-month baseline, attributed to fluctuations in the intrinsic AGN
continuum \citep{ward1983}; simultaneous \textit{Swift} monitoring
revealed correlated variability across the X-ray, UV, and optical
bands, in agreement with X-ray reprocessing by extended regions of the
accretion disc or BLR \citep{mchardy2024}. For NGC~4593,
\citet{Souza2026RM} report multifractality concentrated in the X-ray
bands ($\Delta\alpha = 0.274$ for HX and $0.348$ for SX), with a
strongly right-skewed W2 spectrum ($A = 2.456$) and near-monofractal
B and V bands.
 
The IDRM strategy improves on earlier reverberation-mapping campaigns by incorporating six UV/optical filters, a higher sampling cadence, and substantially more observations, enhancing sensitivity to short inter-band lags. It was first applied to NGC~5548 and later
extended to additional targets \citep{edelson2015space, mchardy2018x,
edelson2019first}, with simultaneous \textit{HST} and ground-based
coverage.

The light curves show strong variability in all UVOT bands
and in HX, with more moderate variability in SX. The IDRM cadence
adequately tracks UV/optical variations, whereas X-ray fluctuations
occur on shorter timescales and may be partially undersampled
(\citealt{edelson2019first}). The fundamental properties of the four
AGNs , redshift, black hole mass, Eddington ratio, and cadence are
summarised in Table~\ref{tab:agn_properties}.
 
\begin{table*}
    \centering
    \caption{Fundamental properties of the selected AGNs, including
    redshift ($z$), black hole mass ($\log M_{\mathrm{BH}}$), Eddington
    ratio ($\dot{m}_{\mathrm{Edd}}$), campaign time span, total
    duration, number of visits, and mean sampling cadence. Redshifts
    are drawn from the SIMBAD database. Black hole masses are adopted
    from the AGN Black Hole Mass Database \citep{2015BentzBlackHole}.
    The Eddington ratio is defined as
    $\dot{m}_{\mathrm{Edd}} = L_{\mathrm{bol}} / L_{\mathrm{Edd}}$.
    Campaign duration is expressed in days, the number of visits
    corresponds to all observations with usable data in at least one
    band, and the mean cadence is derived from the ratio of campaign
    duration to number of visits.}
    \label{tab:agn_properties}
    \begin{tabular}{lcccccc}
        \hline
        Name & $z$ & $\log M_{\text{BH}}$ & $\dot{m}$ & Visits & Duration (days) & Cadence (days)\\
        \hline
        Mrk~509  & 0.0341 & 8.05 & 5\% & 257 & 272.6 & 1.065\\
        NGC~5548 & 0.0163 & 7.72 & 5\% & 291 & 127.6 & 0.440\\
        NGC~4151 & 0.0032 & 7.56 & 1\% & 322 & 69.3  & 0.216\\
        NGC~4593 & 0.0083 & 6.88 & 8\% & 194 & 22.6  & 0.117\\
        \hline
    \end{tabular}
\end{table*}

% ==================================================================
\section{Methods}
\label{sec:method}
% ==================================================================

Information theory, originally formulated by \citet{shannon1948} in the
context of communication systems, provides a natural framework for
characterising statistical dependence in a model-free, nonlinear
fashion. The central quantity is the Shannon entropy,
\begin{equation}
    H(X) = -\int p(x)\log p(x)\,\mathrm{d}x,
    \label{eq:shannon}
\end{equation}
where $p(x)$ is the probability density function of $X$, and $H(X)$ measures the average uncertainty in the random variable $X$ \citep{cover2006}. From this foundation, one constructs the Mutual Information (MI) between two random variables $X$ and $Y$
(Eq.~\ref{eq:mi}).  Crucially, MI vanishes if and only if $X$ and
$Y$ are statistically independent, and captures any form of dependence,
linear or nonlinear, without assuming a specific functional form
\citep{cover2006}.  MI has been widely applied in time-series analysis
across diverse fields, including neuroscience
\citep{kraskov2004estimating, wibral2014transfer}, climatology and
geophysical signals \citep{telesca2011magneto, telesca2011wind}, and
financial markets \citep{Fiedor2014}. The $k$-nearest-neighbour method of
\citet{kraskov2004estimating} is widely adopted for this
purpose and is available via the \textsc{scikit-learn} library.
  Using MI as a lag
estimator, by maximising the MI as a function of a trial time shift,
$\tau$, is a natural generalisation of the CCF that requires no
linearity assumption. 

To characterise temporal dependencies between the multi-band light
curves of our four Seyfert galaxies, we employ a sequential, self-consistent
analysis implemented as a standalone Python pipeline. We estimate the
inter-band time lag by maximising the time-lagged 
MI as a function of trial lag, with uncertainties derived from a
Random Subset Selection bootstrap. The pipeline relies on
\texttt{numpy}, \texttt{scipy}, and \texttt{matplotlib}, with the
MI estimation step performed by the $k$-nearest-neighbour mutual
information estimator \citep{kraskov2004estimating} implemented in
the \textsc{scikit-learn} library.

\subsection{Mutual Information Lag Estimation}
\label{subsec:method_mi}

\subsubsection{Mutual Information as a lag estimator}

The MI between two continuous random variables $X$
and $Y$ is defined as \citep{cover2006}:
\begin{equation}
    \mathrm{MI}(X;Y) = \iint p(x,y)\,\log\frac{p(x,y)}{p(x)\,p(y)}\,dx\,dy,
    \label{eq:mi}
\end{equation}
where $p(x,y)$ is the joint probability density function (PDF) and
$p(x)$, $p(y)$ are the marginal PDFs. Practical computation from
finite data requires a non-parametric estimator of the densities
entering Eq.~(\ref{eq:mi}). Higher MI values indicate a stronger
dependence between the two variables, whereas MI becomes zero if and
only if $X$ and $Y$ are statistically independent. Unlike the Pearson
correlation coefficient, MI captures any statistical dependence, linear
or non-linear, making it sensitive to the full structure of the joint
variability between two AGN Light Curves.

For a pair of light curves, we define the time-lagged MI curve as
\begin{equation}
    \mathrm{MI}(\tau) = \mathrm{MI}\!\left[x_{\mathrm{ref}}(t - \tau),\;
    x_{\mathrm{sub}}(t)\right],
    \label{eq:mi_lag}
\end{equation}
where $x_{\mathrm{ref}}$ is the reference band and $x_{\mathrm{sub}}$
is the subsidiary band. In this convention, a
positive lag $\tau > 0$ corresponds to the reference band
\emph{leading} the subsidiary band, which is the expected sign for
disc reprocessing, in which the UV continuum leads the longer-wavelength
optical response \citep{cackett2007testing}. The lag that maximises
$\mathrm{MI}(\tau)$ is taken as the estimated inter-band time delay,
as that displacement maximises the statistical information shared
between the two series.

The shift is implemented by linear interpolation: for each trial lag
$\tau$, both light curves are first resampled onto a common uniform
time grid spanning their temporal overlap, and the subsidiary series
is then evaluated on that grid after an additional shift of $\tau$
using \texttt{scipy.interpolate.interp1d}. This produces the pair
$\{x_{\mathrm{ref}}(t), x_{\mathrm{sub}}(t+\tau)\}$ on the common grid,
which is then passed to the MI estimator. This manual resampling is
required because the $k$-nearest-neighbour MI estimator has no native
lag parameter and must be evaluated independently for each trial
$\tau$.

\subsubsection{KSG estimator}
\label{subsubsec:ksg}

The PDFs in Eq.~(\ref{eq:mi}) are unknown and must be estimated
from the data. We adopt the Kraskov-Stoegbauer-Grassberger (KSG)
$k$-nearest-neighbour estimator \citep{kraskov2004estimating}, as
implemented in the \texttt{mutual\_info\_regression} function of the
\textsc{scikit-learn} library. For each observation
$i$, the distance $\epsilon_i$ to its $k$-th nearest neighbour is
computed in the joint space $(x_{\mathrm{ref}},\, x_{\mathrm{sub}})$
using the maximum norm. The marginal neighbour counts
$n_x(i)$ and $n_y(i)$ are then defined as the number of points whose
distance from the $i$-th point is strictly smaller than
$\epsilon_i/2$ in the one-dimensional $x_{\mathrm{ref}}$ and
$x_{\mathrm{sub}}$ subspaces, respectively. MI is estimated
as:
\begin{equation}
    \widehat{\mathrm{MI}}(\tau; k) = \psi(k) + \psi(N)
    - \frac{1}{N}\sum_{i=1}^{N}
      \left[\psi\!\left(n_x(i)+1\right) + \psi\!\left(n_y(i)+1\right)\right],
    \label{eq:ksg}
\end{equation}
where $\psi$ is the digamma function, $\psi(x) = \mathrm{d}\ln\Gamma(x)/\mathrm{d}x$,
and $N$ is the number of data points. We write
$\widehat{\mathrm{MI}}(\tau; k)$ explicitly as a function of the time
lag $\tau$, since $x_{\mathrm{sub}}$ is shifted by $\tau$ relative to
$x_{\mathrm{ref}}$ before the joint and marginal neighbour counts are
computed, and of $k$, the number of nearest neighbours used in the
estimator. We use $k = 5$ nearest neighbours, the value recommended by
\citet{kraskov2004estimating} for moderate sample sizes. The KSG
estimator is asymptotically unbiased, requires no bandwidth or binning
choice, and is well suited to the moderately sized, normalised
continuous series produced by the light-curve pre-processing described
below.

\subsubsection{Light-curve pre-processing}
\label{subsubsec:preproc}

Each light curve is sorted by observation time and standardised to
zero mean and unit variance.

Following \citet{edelson2019first}, we adopt the W2 band 
($\lambda_{\rm eff} = 1928$~\AA) as the reference for all inter-band 
lag measurements, as it provides the shortest-wavelength UVOT 
continuum reference with high variability amplitude and dense 
temporal coverage across all four campaigns. For each object, the reference band and
subsidiary light curves are linearly interpolated onto a common
uniform time grid with spacing $\Delta t$ covering their temporal
overlap; grid points at which either interpolant is undefined are
flagged as \texttt{NaN} and excluded via a finite-value mask. Trial
lags for which fewer than 20 valid common points remain are excluded
from the $\mathrm{MI}(\tau)$ curve entirely, as the KSG estimator
is unreliable at such sample sizes.

Because the four campaigns differ substantially in total duration and
mean cadence (Table~\ref{tab:agn_properties}, spanning from
$\sim$23\,d for NGC~4593 to $\sim$273\,d for Mrk~509), the trial-lag
range, lag step, and grid spacing $\Delta t$ are tuned for each object to
match the relevant timescales while keeping the computational cost of
the bootstrap (Section~\ref{subsubsec:rss}) tractable. The grid spacing $\Delta t$ is matched to the mean sampling cadence of each campaign (Table~\ref{tab:agn_properties}), while $\tau_{\max}$ is scaled to between $5\%$ and $18\%$ of the total 
campaign duration (Table~\ref{tab:lag_params}), reflecting the 
substantially different baselines and expected lag scales across 
the four objects, so that the lag-scan parameters reflect the temporal sampling and baseline available for each object.

\begin{table}
    \centering
    \caption{Lag-scan parameters adopted for each object: maximum
    trial lag $\tau_{\max}$, lag step $\delta\tau$, common-grid
    spacing $\Delta t$, number of trial lags $N_{\tau}$, and the
    fraction of the total campaign duration $\tau_{\max}/T$ covered
    by the lag scan.}
    \label{tab:lag_params}
    \begin{tabular}{lccccc}
        \hline
        Object & $\tau_{\max}$ (d) & $\delta\tau$ (d) & $\Delta t$ (d) & 
        $N_{\tau}$ & $\tau_{\max}/T$ \\
        \hline
        NGC~4151 & 6  & 0.15 & 0.20 & 81  & 0.087 \\
        NGC~5548 & 10 & 0.15 & 0.25 & 134 & 0.078 \\
        NGC~4593 & 4  & 0.10 & 0.15 & 81  & 0.177 \\
        Mrk~509  & 15 & 0.25 & 0.50 & 121 & 0.055 \\
        \hline
    \end{tabular}
\end{table}

For each object, the trial-lag grid spans $\tau \in
[-\tau_{\max}, +\tau_{\max}]$ in steps of $\delta\tau$. The peak of
$\mathrm{MI}(\tau)$ is located by fitting a second-order polynomial to
all points with $\mathrm{MI} \geq 0.70\,\mathrm{MI}_{\max}$:
\begin{equation}
    \mathrm{MI}(\tau) \approx a\tau^2 + b\tau + c,
    \qquad \hat{\tau}_{\mathrm{peak}} = -\frac{b}{2a},
    \label{eq:parabola}
\end{equation}
providing a sub-step peak estimate.

\subsubsection{Random Subset Selection bootstrap}
\label{subsubsec:rss}

To quantify the uncertainty on $\hat{\tau}_{\mathrm{peak}}$, we apply
a Random Subset Selection (RSS) bootstrap \citep{efron1979bootstrap}:
in each of $N_{\mathrm{boot}} = 150$ iterations, a fraction $f = 0.80$
of the data points is drawn \emph{without replacement} from each light
curve independently, and the full analysis (interpolation,
normalisation, MI curve computation, and peak detection) is repeated
on the subset. Sampling without replacement preserves the temporal
ordering of the selected points and avoids artificially duplicated
observations that could bias the density estimate.

The 150 bootstrap peak values $\{\hat{\tau}_i\}$ characterise the lag
uncertainty. We report:
\begin{align}
    \hat{\tau}_{\mathrm{med}} &= \mathrm{median}\!\left(\{\hat{\tau}_i\}\right),
    \label{eq:boot_med} \\
    \sigma^{+} &= P_{84}\!\left(\{\hat{\tau}_i\}\right) - \hat{\tau}_{\mathrm{med}},
    \label{eq:boot_hi} \\
    \sigma^{-} &= \hat{\tau}_{\mathrm{med}} - P_{16}\!\left(\{\hat{\tau}_i\}\right),
    \label{eq:boot_lo}
\end{align}
where $P_{16}$ and $P_{84}$ denote the 16th and 84th percentiles of
the bootstrap distribution, corresponding to the $1\sigma$ interval
of a Gaussian distribution. The estimated lag and its asymmetric
uncertainty are then reported as
$\hat{\tau}_{\mathrm{med}}\,^{+\sigma^{+}}_{-\sigma^{-}}$.

The output of the full procedure consists of two diagnostic plots per
band pair: the normalised MI lag curve $\mathrm{MI}(\tau)$
(Fig.~\ref{fig:ngc5548_kernelmi}) and the histogram of bootstrap peak
values (Fig.~\ref{fig:ngc5548_bootstrap}).  In a well-behaved case
such as NGC~5548/B ($\lambda = 4392$\,\AA), the MI curve displays a
single, well-isolated peak and the bootstrap distribution is
approximately Gaussian and narrow ($\sigma^{\pm} \lesssim 0.17$\,d).
Conversely, in bands with weak physical coupling (e.g.\ the X-ray
bands HX and SX), the MI curve is broad and multi-peaked, and the
bootstrap histogram is correspondingly wide and skewed, signalling
that no reliable lag can be extracted.

% ==================================================================
\section{Results and Discussion}
\label{sec:results}
% ==================================================================

\begin{figure*}
\centering
\includegraphics[width=0.32\textwidth]{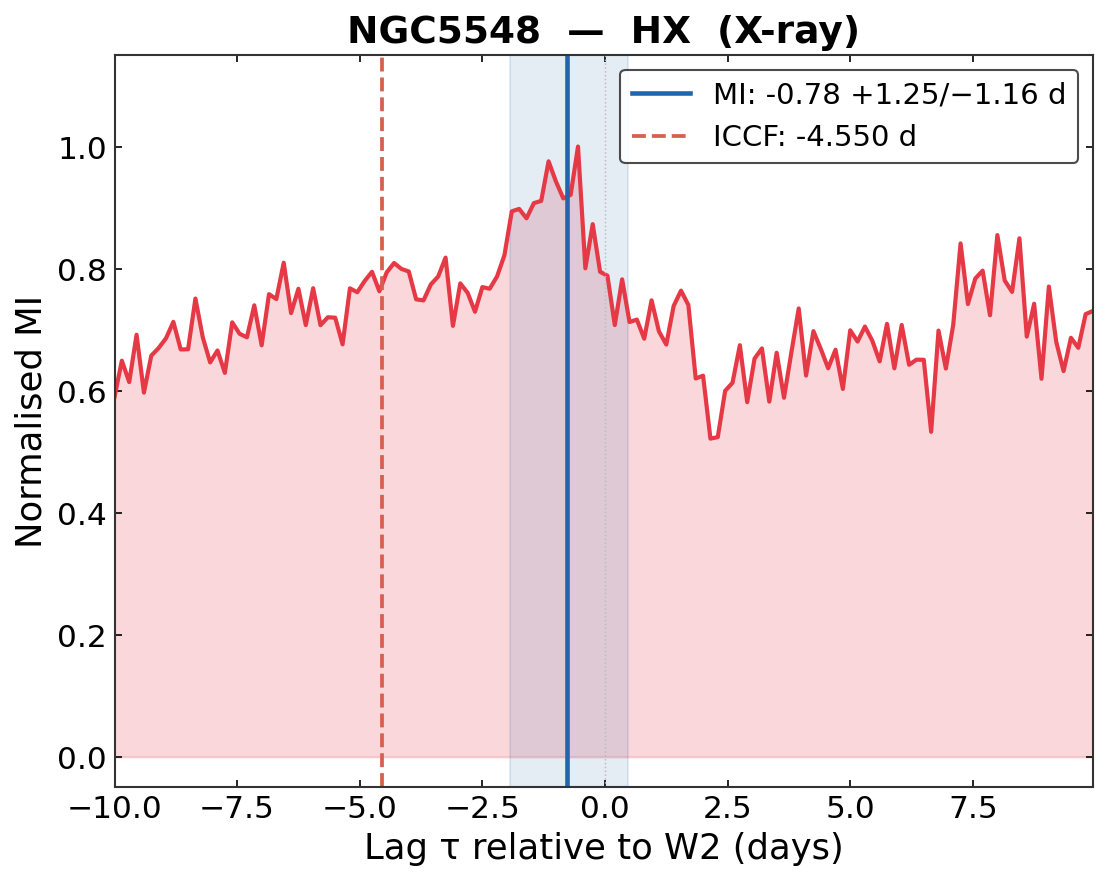}\hfill
\includegraphics[width=0.32\textwidth]{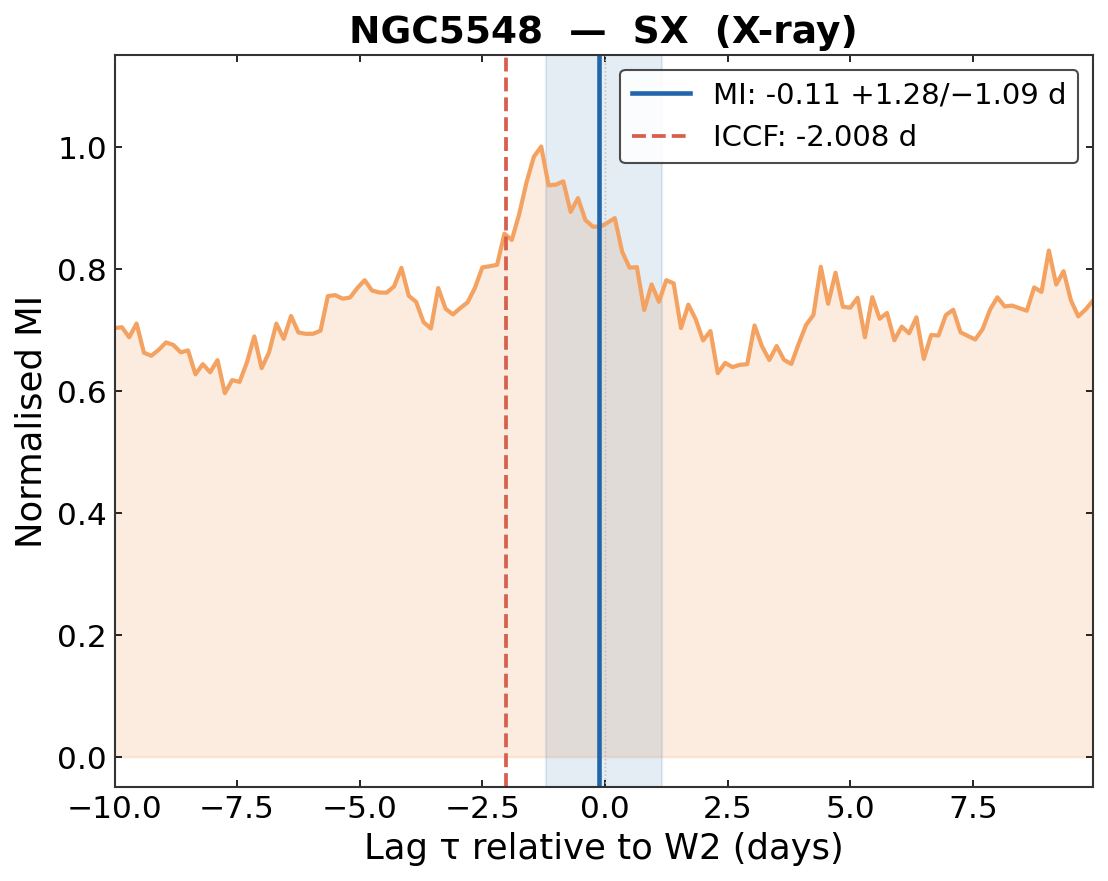}\hfill
\includegraphics[width=0.32\textwidth]{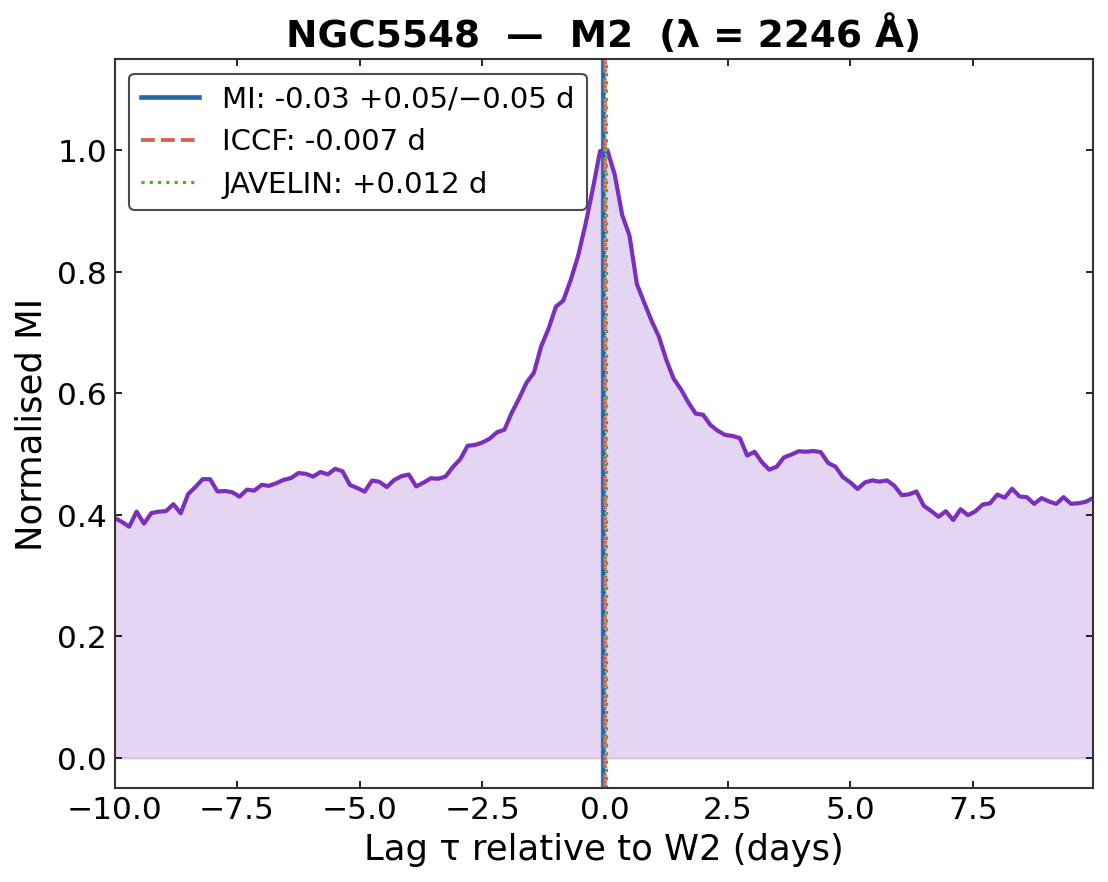}\\[4pt]
\includegraphics[width=0.32\textwidth]{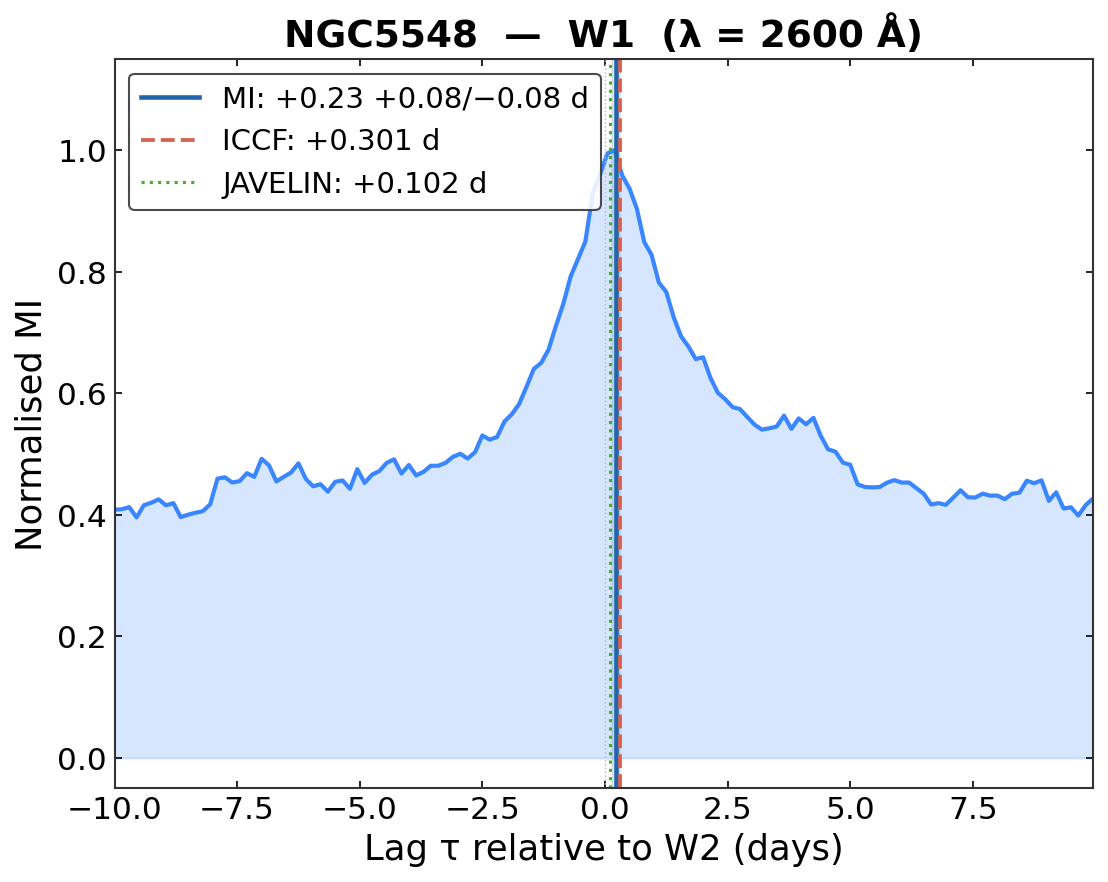}\hfill
\includegraphics[width=0.32\textwidth]{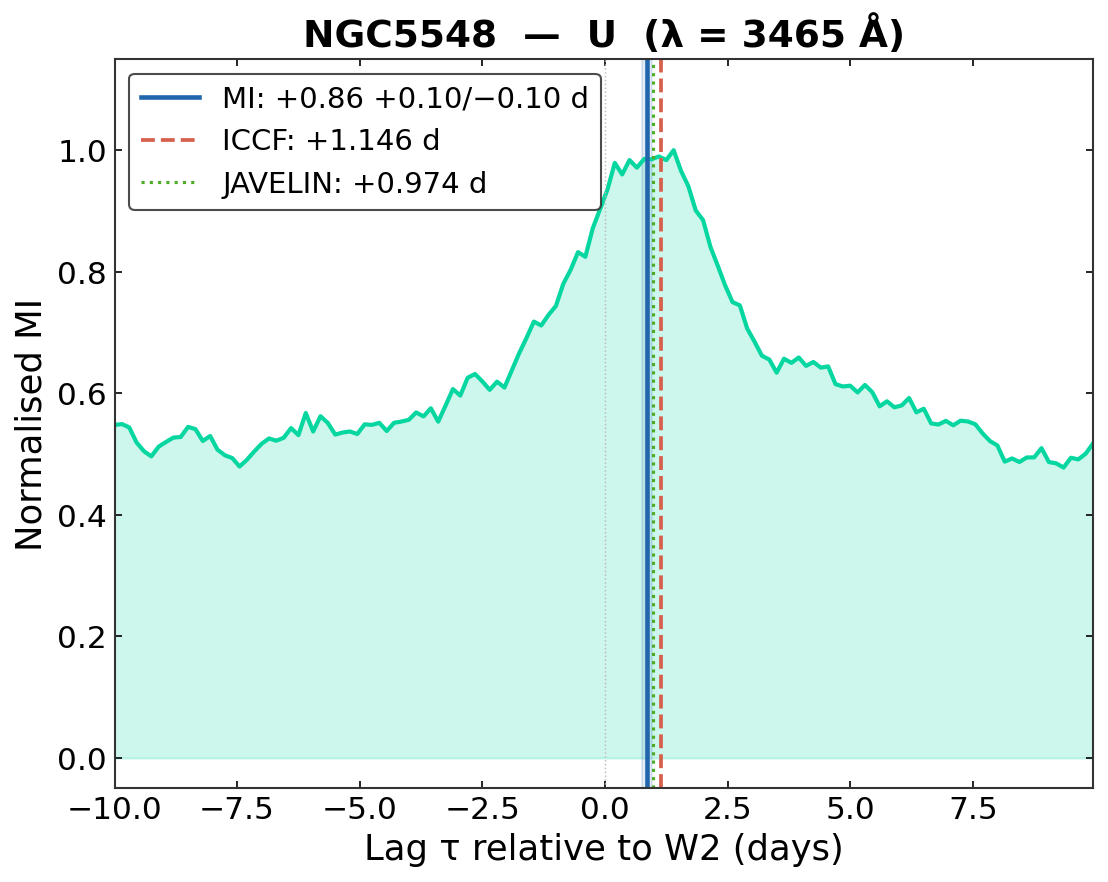}\hfill
\includegraphics[width=0.32\textwidth]{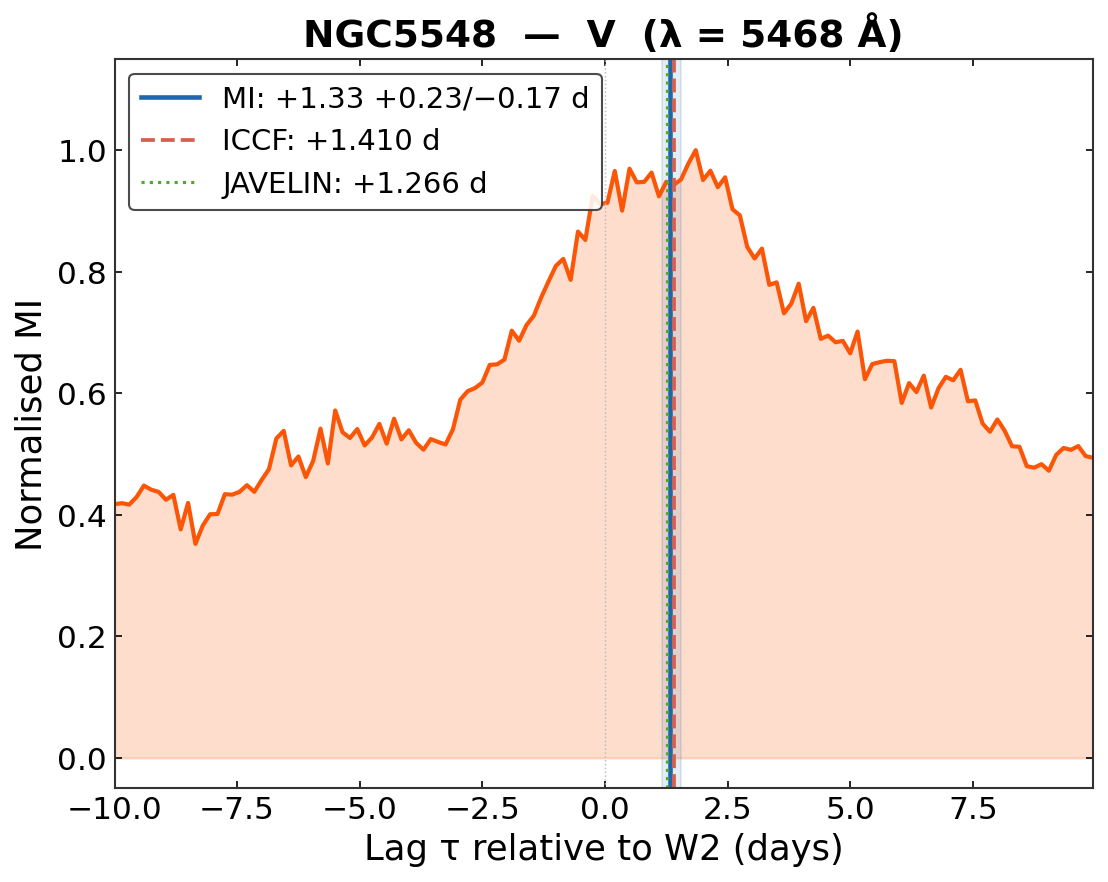}\\[4pt]
\includegraphics[width=0.32\textwidth]{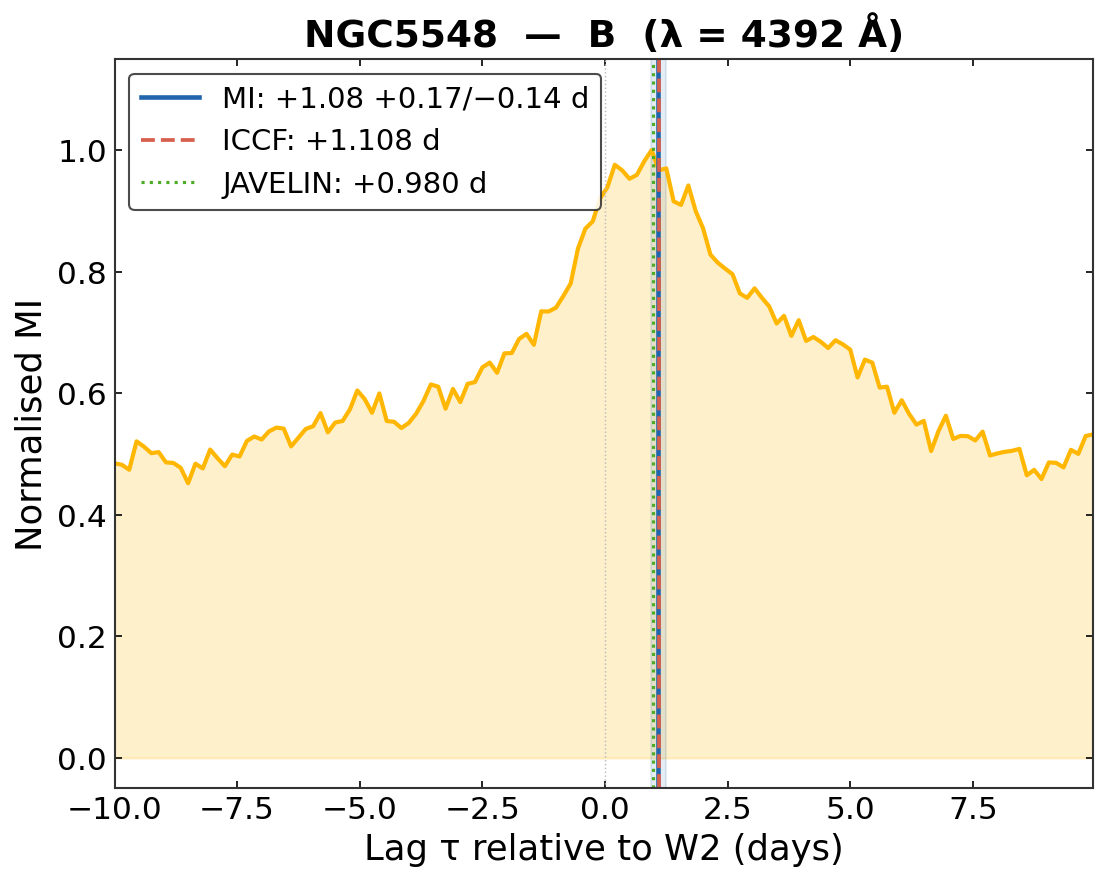}
\caption{Normalised MI lag curves for all seven bands of NGC~5548,
  relative to the W2 reference band. The solid blue vertical line
  marks the bootstrap median lag $\hat{\tau}_{\mathrm{med}}$; the
  blue shaded region spans the 16th to 84th percentile interval. The
  red dashed and green dotted lines show the ICCF and JAVELIN
  reference lags of \citet{edelson2019first}, respectively, where
  available. The UV bands (M2 through U) display a single well-defined
  peak consistent with disc reprocessing; the X-ray bands (HX, SX)
  show broad, multi-peaked curves indicative of weak X-ray - UV
  coupling.}
\label{fig:ngc5548_kernelmi}
\end{figure*}

\begin{figure*}
\centering
\includegraphics[width=0.32\textwidth]{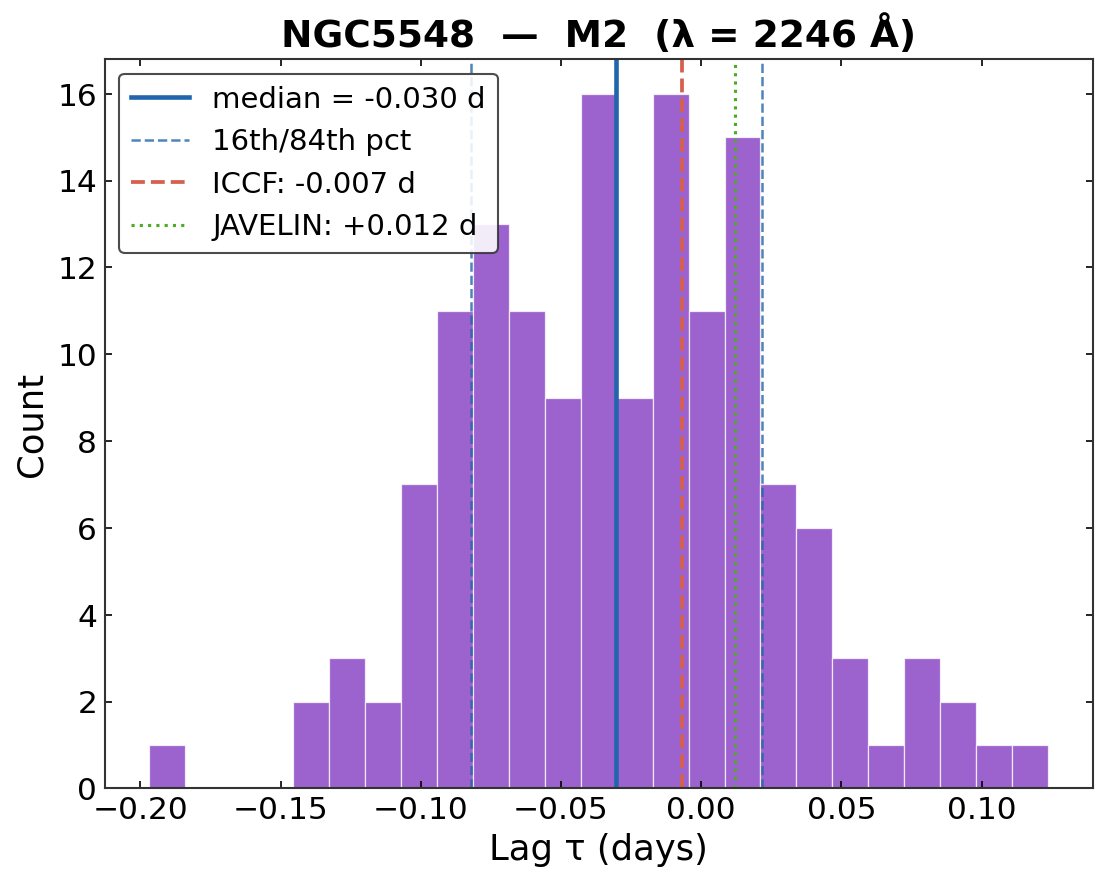}\hfill
\includegraphics[width=0.32\textwidth]{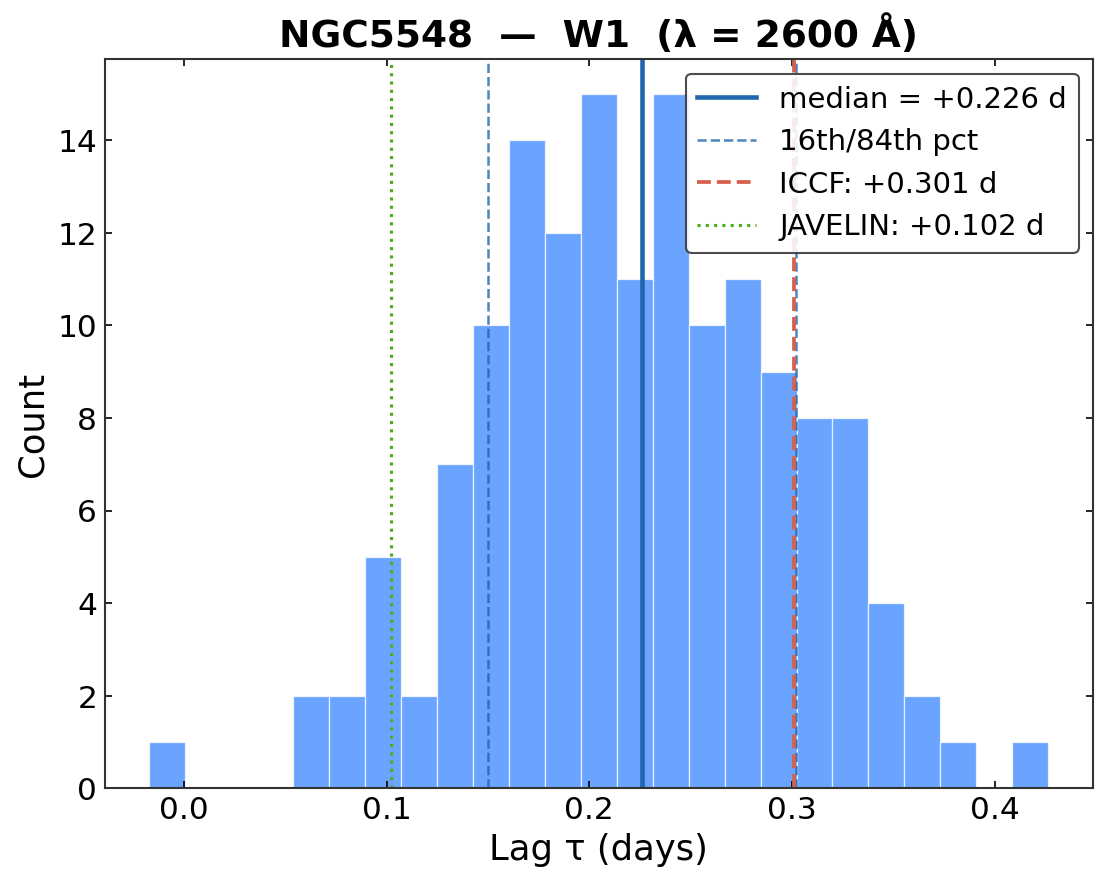}\hfill
\includegraphics[width=0.32\textwidth]{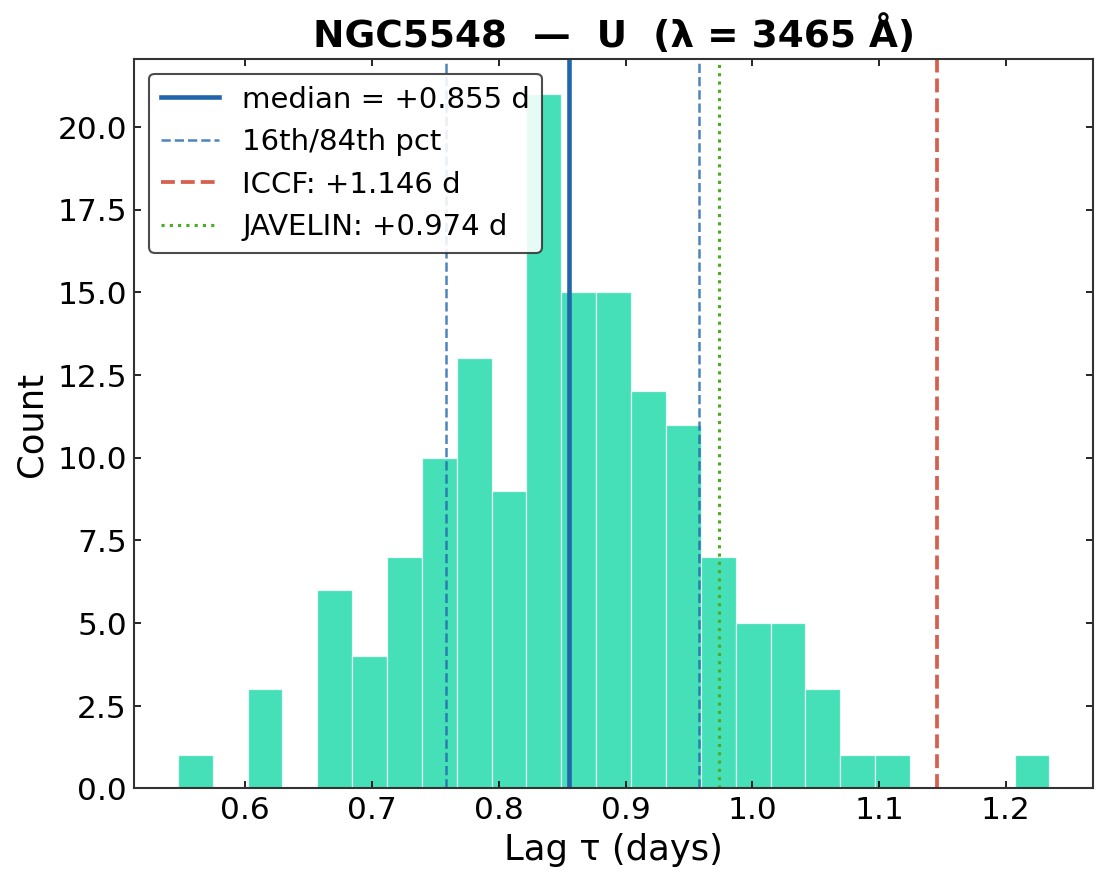}\\[4pt]
\includegraphics[width=0.32\textwidth]{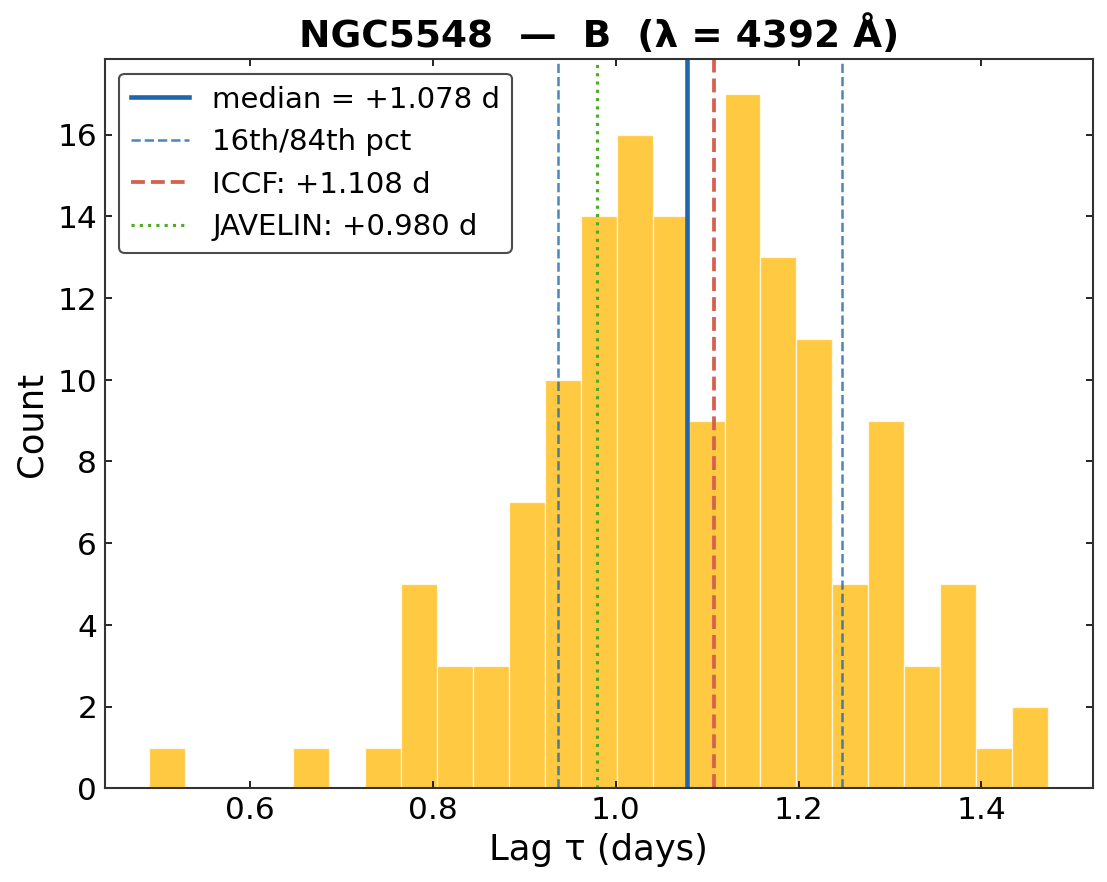}\hfill
\includegraphics[width=0.32\textwidth]{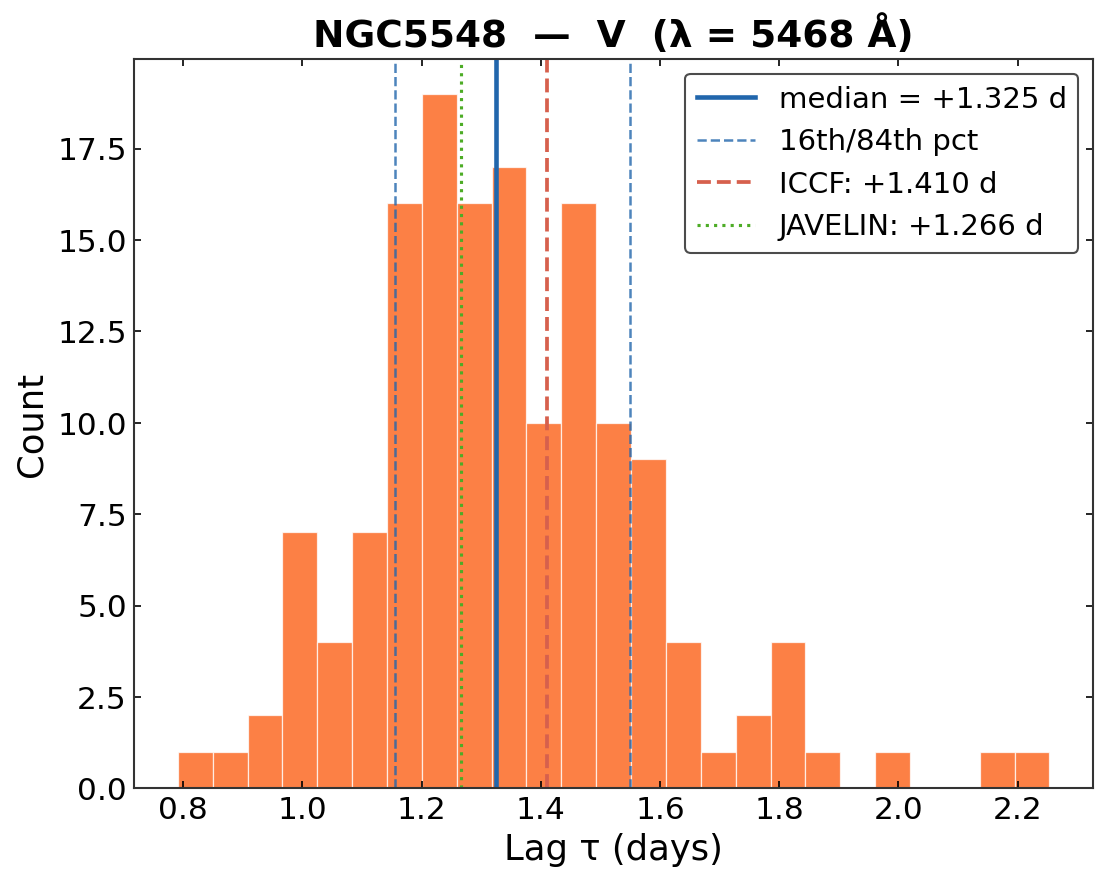}\hfill
\includegraphics[width=0.32\textwidth]{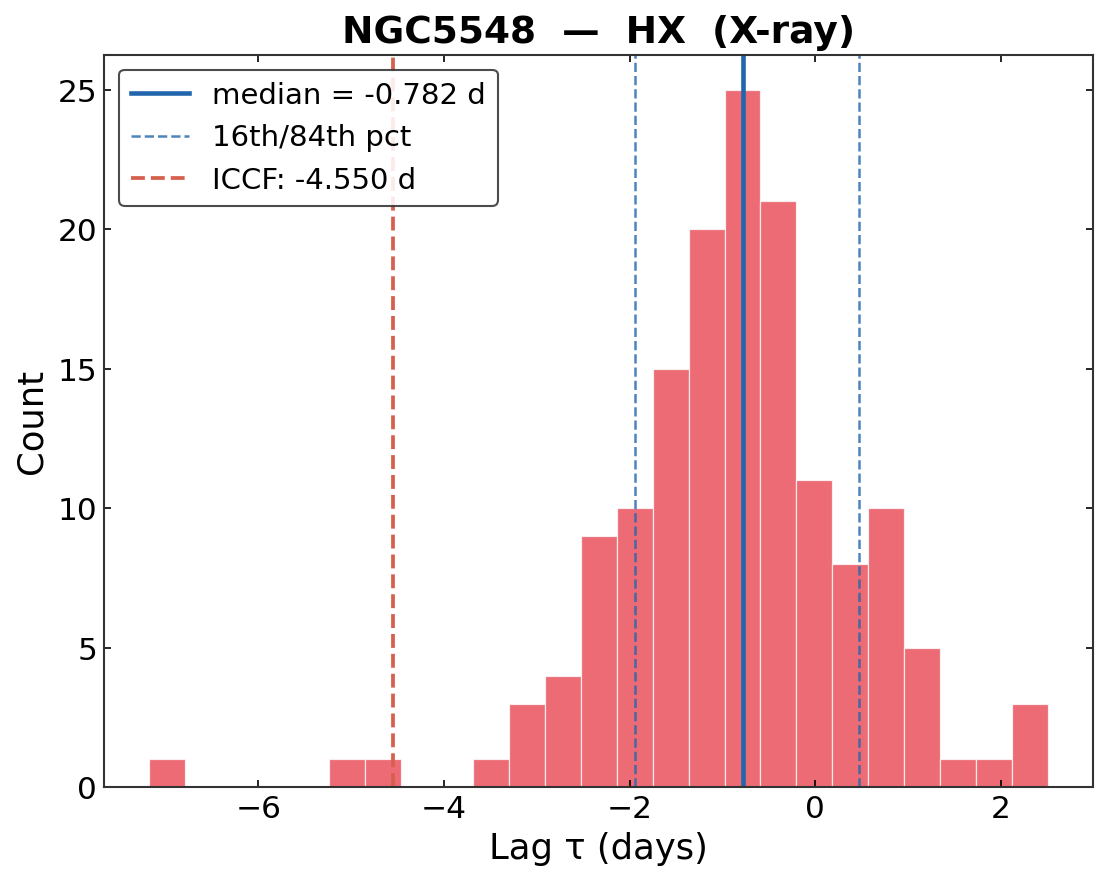}\\[4pt]
\includegraphics[width=0.32\textwidth]{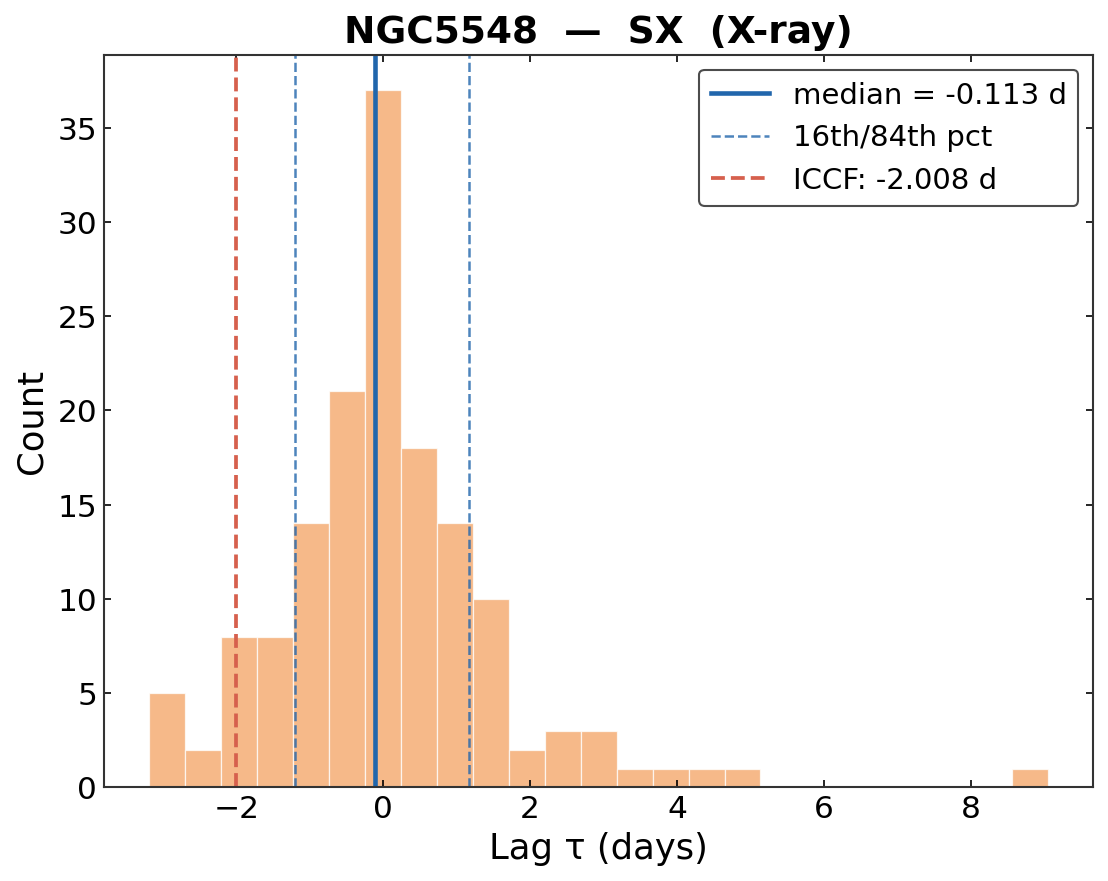}
\caption{Bootstrap lag distributions ($N_{\mathrm{boot}}=150$, RSS
  $f=0.80$) for each band of NGC~5548. The solid blue vertical line
  marks the bootstrap median $\hat{\tau}_{\mathrm{med}}$; dashed blue
  lines indicate the 16th and 84th percentiles. The red dashed and
  green dotted lines show the ICCF and JAVELIN reference lags of
  \citet{edelson2019first}, respectively, where available. UV bands
  yield compact, approximately Gaussian distributions; the X-ray bands
  HX and SX produce broad, skewed histograms consistent with an
  unresolvable lag.}
\label{fig:ngc5548_bootstrap}
\end{figure*}

\begin{figure*}
\centering
\includegraphics[width=0.49\textwidth]{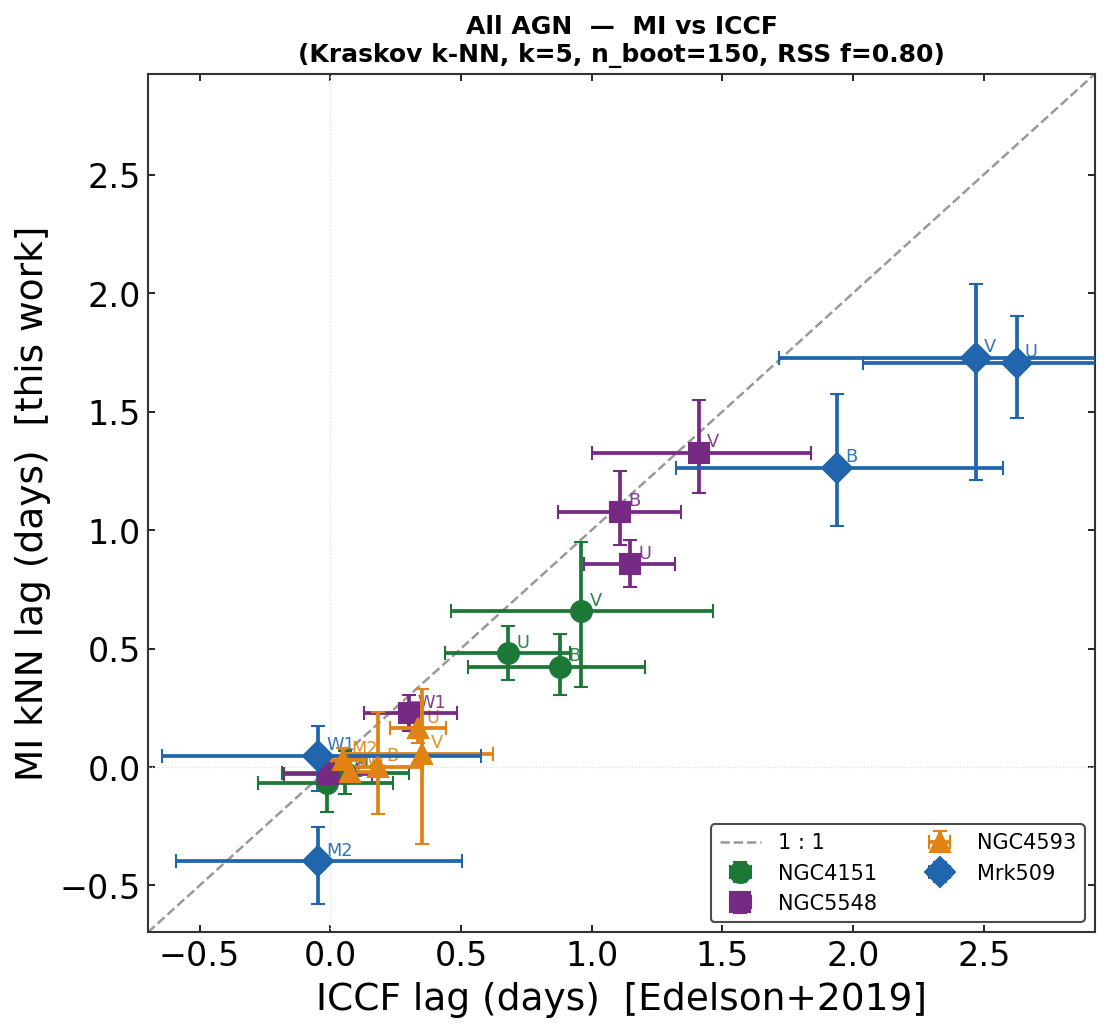}\hfill
\includegraphics[width=0.49\textwidth]{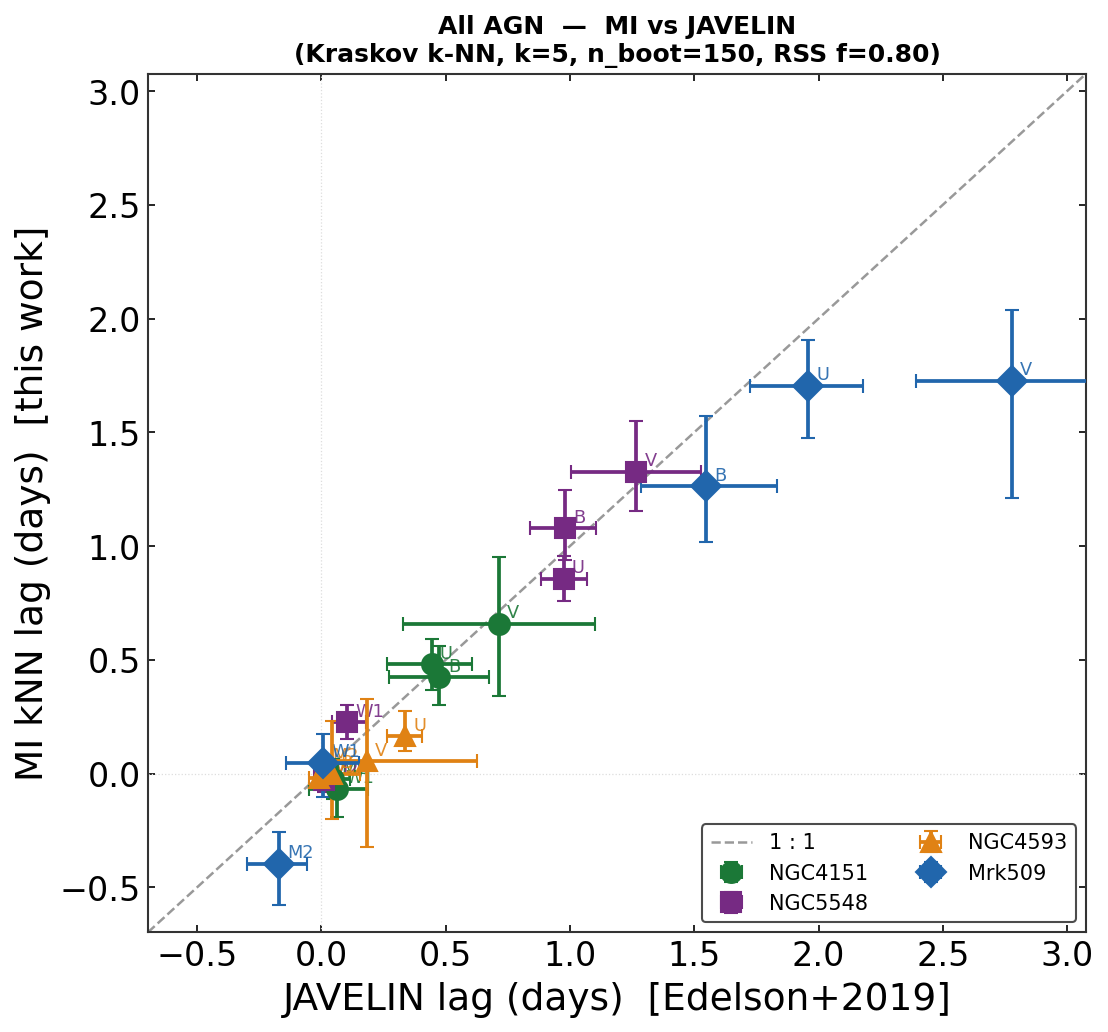}
\caption{MI lag versus ICCF (left) and JAVELIN (right)
  reference lags for all four AGNs. Each panel shows the MI bootstrap
  median lag (with $1\sigma$ RSS uncertainties) against the reference
  lag of \citet{edelson2019first}, relative to the W2 reference band.
  The dashed line marks the 1:1 relation.}
\label{fig:four_panels}
\end{figure*}

The KSG MI inter-band time lags for all four AGNs are listed in
Table~\ref{tab:kernelmi_lags}. Below we discuss the results source by
source (Sections~\ref{subsec:res_5548},\ref{subsec:res_mrk509}) and
then synthesise the band-level behaviour across the full sample
(Section~\ref{subsec:byband}). The MI lag curves and bootstrap
distributions for NGC~5548 are shown as representative examples in
Figs.~\ref{fig:ngc5548_kernelmi} and \ref{fig:ngc5548_bootstrap};
equivalent figures for the remaining sources are provided as online 
supplementary material. The comparison between MI lags and the 
ICCF/JAVELIN reference values of \citet{edelson2019first} for all 
four AGNs is shown in Fig.~\ref{fig:four_panels}.

\begin{table*}
    \centering
    \caption{KSG MI inter-band time lags relative to the W2 (UVW2)
             reference band for the four AGNs in the sample.
             $\hat{\tau}_{\mathrm{med}}$ is the bootstrap median lag,
             with $\sigma^{-}$ and $\sigma^{+}$ denoting the
             $1\sigma$ asymmetric uncertainties derived from the 16th
             and 84th percentiles of 150 RSS bootstrap realisations.
             $\tau_{\mathrm{ICCF}}$ and $\tau_{\mathrm{JAV}}$ are the
             ICCF and JAVELIN literature lags from
             \citet{edelson2019first} (JAVELIN not available for X-ray
             bands). $\Delta\alpha$ is the multifractal spectrum width
             for each band, taken from \citet{Souza2026RM}.}
    \label{tab:kernelmi_lags}
    \setlength{\tabcolsep}{6pt}
    \begin{tabular}{llrrrrrr}
        \hline
        Object & Band
               & $\hat{\tau}_{\mathrm{med}}$ (d)
               & $\sigma^{-}$ (d)
               & $\sigma^{+}$ (d)
               & $\tau_{\mathrm{ICCF}}$ (d)
               & $\tau_{\mathrm{JAV}}$ (d)
               & $\Delta\alpha$ \\
        \hline
        \multirow{7}{*}{NGC~5548}
          & M2  & $-0.030$ & $0.052$ & $0.052$ & $-0.007$ & $+0.012$ & $0.244$ \\
          & W1  & $+0.226$ & $0.076$ & $0.076$ & $+0.301$ & $+0.102$ & $0.186$ \\
          & U   & $+0.855$ & $0.097$ & $0.103$ & $+1.146$ & $+0.974$ & $0.153$ \\
          & B   & $+1.078$ & $0.141$ & $0.170$ & $+1.108$ & $+0.980$ & $0.085$ \\
          & V   & $+1.325$ & $0.169$ & $0.225$ & $+1.410$ & $+1.266$ & $0.038$ \\
          & HX  & $-0.782$ & $1.157$ & $1.250$ & $-4.550$ & - & $0.293$ \\
          & SX  & $-0.113$ & $1.094$ & $1.283$ & $-2.008$ & - & $0.681$ \\
        \hline
        \multirow{7}{*}{NGC~4151}
          & M2  & $-0.025$ & $0.088$ & $0.091$ & $+0.055$ & $+0.045$ & $0.017$ \\
          & W1  & $-0.066$ & $0.109$ & $0.124$ & $-0.011$ & $+0.064$ & $0.017$ \\
          & U   & $+0.480$ & $0.114$ & $0.113$ & $+0.679$ & $+0.443$ & $0.018$ \\
          & B   & $+0.423$ & $0.120$ & $0.139$ & $+0.877$ & $+0.475$ & $0.008$ \\
          & V   & $+0.658$ & $0.318$ & $0.294$ & $+0.960$ & $+0.714$ & $0.003$ \\
          & HX  & $+0.508$ & $0.872$ & $1.163$ & $-3.324$ & - & $0.542$ \\
          & SX  & $-2.715$ & $1.928$ & $1.653$ & $-2.408$ & - & $0.040$ \\
        \hline
        \multirow{7}{*}{NGC~4593}
          & M2  & $+0.030$ & $0.034$ & $0.050$ & $+0.048$ & $+0.009$ & $0.120$ \\
          & W1  & $-0.020$ & $0.043$ & $0.059$ & $+0.077$ & $-0.010$ & $0.077$ \\
          & U   & $+0.163$ & $0.063$ & $0.111$ & $+0.337$ & $+0.337$ & $0.054$ \\
          & B   & $+0.000$ & $0.200$ & $0.231$ & $+0.182$ & $+0.041$ & $0.013$ \\
          & V   & $+0.054$ & $0.379$ & $0.275$ & $+0.351$ & $+0.182$ & $0.004$ \\
          & HX  & $-0.531$ & $0.490$ & $0.350$ & $-0.602$ & - & $0.274$ \\
          & SX  & $-0.405$ & $0.702$ & $0.273$ & $-0.538$ & - & $0.348$ \\
        \hline
        \multirow{7}{*}{Mrk~509}
          & M2  & $-0.398$ & $0.178$ & $0.143$ & $-0.047$ & $-0.172$ & $0.194$ \\
          & W1  & $+0.046$ & $0.148$ & $0.129$ & $-0.047$ & $+0.005$ & $0.144$ \\
          & U   & $+1.705$ & $0.230$ & $0.200$ & $+2.626$ & $+1.959$ & $0.123$ \\
          & B   & $+1.263$ & $0.245$ & $0.310$ & $+1.937$ & $+1.548$ & $0.089$ \\
          & V   & $+1.727$ & $0.517$ & $0.310$ & $+2.469$ & $+2.776$ & $0.056$ \\
          & HX  & $-0.915$ & $1.535$ & $1.183$ & $+4.941$ & - & $0.216$ \\
          & SX  & $+0.824$ & $0.751$ & $0.848$ & $+2.332$ & - & $0.321$ \\
        \hline
    \end{tabular}
\end{table*}

\subsection{NGC~5548}
\label{subsec:res_5548}

The MI lag curves and bootstrap distributions for NGC~5548 are shown
in Figs.~\ref{fig:ngc5548_kernelmi} and \ref{fig:ngc5548_bootstrap}.
In the UV bands M2, W1, and U the MI returns
$-0.030^{+0.052}_{-0.052}$, $+0.226^{+0.076}_{-0.076}$, and
$+0.855^{+0.103}_{-0.097}$\,d, respectively, against ICCF values of $-0.007$,
$+0.301$, and $+1.146$\,d \citep{edelson2019first}, giving
$|\Delta\tau| = 0.02$, $0.08$, and $0.29$\,d respectively. The
JAVELIN lags ($+0.012$, $+0.102$, $+0.974$\,d) bracket these
estimates from above in the near-UV and are consistent within the
bootstrap uncertainties. In the optical, B returns
$+1.078^{+0.170}_{-0.141}$\,d against $+1.108$\,d (ICCF,
$|\Delta\tau| = 0.03$\,d; JAVELIN $+0.980$\,d), and V returns
$+1.325^{+0.225}_{-0.169}$\,d against $+1.410$\,d (ICCF,
$|\Delta\tau| = 0.09$\,d; JAVELIN $+1.266$\,d). The bootstrap
histograms for these bands are compact and approximately Gaussian,
confirming a well-constrained peak.

In the X-ray bands the MI lags depart from the ICCF reference. For
HX the MI returns $-0.782^{+1.250}_{-1.157}$\,d against $-4.550$\,d
(ICCF, $|\Delta\tau| = 3.77$\,d); for SX, $-0.113^{+1.283}_{-1.094}$\,d
against $-2.008$\,d ($|\Delta\tau| = 1.90$\,d). Both bootstrap
distributions are wide and skewed (Fig.~\ref{fig:ngc5548_bootstrap}),
consistent with the broad, multi-peaked MI curves in
Fig.~\ref{fig:ngc5548_kernelmi}, and the ICCF lags lie well outside
the RSS uncertainty intervals in both cases.

Physically, the close UV/optical agreement between MI, ICCF, and JAVELIN in the near-UV (M2, W1) supports the disc-reprocessing picture, in which continuum reprocessing by the accretion disc drives a wavelength-ordered lag sequence \citep{cackett2007testing}. The growing $|\Delta\tau|$ toward U indicates an increasing diffuse continuum/BLR contribution to the observed lag \citep{cackett2018accretion}, while the failure in HX/SX reflects the intrinsically weak X-ray/UV coupling reported for this source by \citet{edelson2019first}, rather than a limitation of the MI estimator itself.

\subsection{NGC~4151}
\label{subsec:res_4151}

For NGC~4151 (Table \ref{tab:kernelmi_lags}), the MI lags in
M2, W1, and U are $-0.025^{+0.091}_{-0.088}$,
$-0.066^{+0.124}_{-0.109}$, and $+0.480^{+0.113}_{-0.114}$\,d,
against ICCF values of $+0.055$, $-0.011$, and $+0.679$\,d
\citep{edelson2019first}, giving $|\Delta\tau| = 0.08$, $0.06$, and
$0.20$\,d. The corresponding JAVELIN lags ($+0.045$, $+0.064$,
$+0.443$\,d) are compatible within the bootstrap uncertainties. The U
band provides the tightest three-way agreement in this source.
For B the MI returns $+0.423^{+0.139}_{-0.120}$\,d against $+0.877$\,d
(ICCF, $|\Delta\tau| = 0.45$\,d; JAVELIN $+0.475$\,d), and for V
$+0.658^{+0.294}_{-0.318}$\,d against $+0.960$\,d (ICCF,
$|\Delta\tau| = 0.30$\,d; JAVELIN $+0.714$\,d).

In the X-ray bands, HX returns $+0.508^{+1.163}_{-0.872}$\,d against
$-3.324$\,d (ICCF, $|\Delta\tau| = 3.83$\,d, opposite sign). The SX
band gives $-2.715^{+1.653}_{-1.928}$\,d against $-2.408$\,d
($|\Delta\tau| = 0.31$\,d). Both X-ray bootstrap distributions are
broad, with the HX histogram spanning several days
(Table~\ref{tab:kernelmi_lags}).

The tight three-way agreement in U, alongside the shorter MI lags in B and V relative to ICCF, is consistent with a disc-dominated UV/optical response with a modest diffuse continuum contribution superposed at longer wavelengths \citep{cackett2018accretion}. The large, sign-discordant X-ray offsets echo the weak physical X-ray/UV connection found by \citet{edelson2019first} for the full sample, suggesting the corona-to-disc reprocessing channel is not the dominant driver of the observed X-ray variability in this source.

\subsection{NGC~4593}
\label{subsec:res_4593}

For NGC~4593 (Table \ref{tab:kernelmi_lags}), the MI lags in
the UV/optical bands are $+0.030^{+0.050}_{-0.034}$ (M2),
$-0.020^{+0.059}_{-0.043}$ (W1), $+0.163^{+0.111}_{-0.063}$ (U),
$+0.000^{+0.231}_{-0.200}$ (B), and $+0.054^{+0.275}_{-0.379}$\,d
(V). The corresponding ICCF values are $+0.048$, $+0.077$, $+0.337$,
$+0.182$, and $+0.351$\,d \citep{edelson2019first}. The M2 band
shows excellent agreement ($|\Delta\tau| = 0.02$\,d), and W1, B, and
V also agree within the bootstrap uncertainties. The JAVELIN lags
($+0.009$, $-0.010$, $+0.337$, $+0.041$, $+0.182$\,d) are broadly
consistent with the MI estimates. The U band shows the largest
UV/optical discrepancy in this source ($|\Delta\tau| = 0.17$\,d).

For the X-ray bands the MI returns $-0.531^{+0.350}_{-0.490}$\,d
(HX) and $-0.405^{+0.273}_{-0.702}$\,d (SX), against ICCF values of
$-0.602$ and $-0.538$\,d ($|\Delta\tau| = 0.07$ and $0.13$\,d). This
is the only source in the sample for which the MI X-ray lags fall
within one $\sigma$ of the ICCF, although the bootstrap distributions
remain broad relative to those of the UV bands.

NGC~4593 is the only source for which the X-ray lags also agree with the ICCF, alongside the tightest UV/optical baseline consistency. This is compatible with the correlated X-ray/UV/optical variability reported by \citet{cackett2018accretion} for this object, pointing to a cleaner, more centrally concentrated reprocessing geometry with comparatively little diffuse BLR contamination, in contrast to the more extended reprocessing regions inferred for the other three AGNs.

\subsection{Mrk~509}
\label{subsec:res_mrk509}

For Mrk~509 (Table \ref{tab:kernelmi_lags}), the MI lags in the
UV/optical bands are $-0.398^{+0.143}_{-0.178}$ (M2),
$+0.046^{+0.129}_{-0.148}$ (W1), $+1.705^{+0.200}_{-0.230}$ (U),
$+1.263^{+0.310}_{-0.245}$ (B), and $+1.727^{+0.310}_{-0.517}$\,d
(V), against ICCF values of $-0.047$, $-0.047$, $+2.626$, $+1.937$,
and $+2.469$\,d \citep{edelson2019first}. The MI recovers the correct
sign for all five bands. W1 shows the closest agreement
($|\Delta\tau| = 0.09$\,d; JAVELIN $+0.005$\,d). The U, B, and V
lags are systematically shorter than both ICCF and JAVELIN reference
values ($|\Delta\tau| = 0.92$, $0.67$, and $0.74$\,d relative to
ICCF), suggesting that for this source the MI peak identifies the
leading edge of the reprocessing response rather than the centroid.

In the X-ray bands, HX returns $-0.915^{+1.183}_{-1.535}$\,d against
$+4.941$\,d (ICCF, $|\Delta\tau| = 5.86$\,d, opposite sign), the
largest discrepancy in the full sample. SX gives
$+0.824^{+0.848}_{-0.751}$\,d against $+2.332$\,d
($|\Delta\tau| = 1.51$\,d; Table~\ref{tab:kernelmi_lags}).

The systematically shorter MI lags in U, B, and V relative to ICCF and JAVELIN mirror the diffuse continuum excess identified by \citet{edelson2019first}, who report U-band lags a factor of $\sim$2.2 above thin-disc predictions due to BLR diffuse continuum emission. Because MI traces the peak rather than the lag-weighted centroid, it is comparatively insensitive to this extended tail, recovering a delay closer to the direct disc response than to the diffuse-continuum-inflated ICCF/JAVELIN values.

\subsection{Band behaviour across the sample}
\label{subsec:byband}

Considering all four AGNs jointly, the M2 band shows
$|\Delta\tau| = 0.02$, $0.08$, $0.02$, and $0.35$\,d for NGC~5548,
NGC~4151, NGC~4593, and Mrk~509 respectively, with the MI recovering
the correct sign in all cases and the JAVELIN lags consistent within
uncertainties. The W1 band yields $|\Delta\tau| = 0.08$, $0.06$,
$0.10$, and $0.09$\,d, the most uniformly small discrepancies in the
sample and consistently the best-constrained bootstrap distributions.
The U band shows good agreement for NGC~5548, NGC~4151, and NGC~4593
($|\Delta\tau| \leq 0.29$\,d) but the MI is shorter than the ICCF for
Mrk~509 ($|\Delta\tau| = 0.92$\,d). The B band discrepancy grows
across sources: $0.03$, $0.45$, $0.18$, and $0.67$\,d. The V band
yields $|\Delta\tau| = 0.09$, $0.30$, $0.30$, and $0.74$\,d, ranging
from close agreement in NGC~5548 to the largest V-band discrepancy in
Mrk~509.

The X-ray bands HX and SX show the largest and least consistent
discrepancies across all sources. For HX: $3.77$, $3.83$, $0.07$, and
$5.86$\,d; for SX: $1.90$, $0.31$, $0.13$, and $1.51$\,d.
NGC~4593 is the single exception where the MI X-ray lags agree with
the ICCF within the bootstrap uncertainties. In two of the four AGNs
(NGC~5548 and Mrk~509/HX) the MI and ICCF X-ray lags differ in sign.
\citet{edelson2019first} report $r_{\rm max} < 0.75$ for all eight
X-ray/UV pairs across the sample, versus $r_{\rm max} > 0.8$ for all
twenty UV/optical pairs.

Across the full sample, the MI lag is smaller than the ICCF lag in 19 of the 20 UV/optical band pairs (the only exception being Mrk\,509/W1; see left panel of Fig.~\ref{fig:four_panels}), pointing to a structural rather than source-specific effect. ICCF reports the \emph{centroid} of the cross-correlation function, which averages over the full width of the response; if the transfer function has an extended tail toward longer delays, as expected from diffuse BLR/continuum reprocessing \citep{cackett2021reverberation}, this tail pulls the centroid toward larger $\tau$. The MI lag, instead, marks the \emph{peak} of MI($\tau$), the single delay with the most shared information, so it is not pulled by that tail and stays closer to the shortest, dominant lag. This difference between an averaging estimator and a peak-seeking one naturally makes MI lags shorter whenever the response is skewed.

The near-uniform excellence of M2 and W1 across all sources reflects their proximity to the disc continuum with minimal diffuse contamination, while the growing B/V discrepancy traces the increasing diffuse-continuum/BLR contribution at longer wavelengths \citep{cackett2018accretion}. The X-ray bands, showing the largest and most inconsistent offsets,
corroborate the weak X-ray/UV coupling
($r_{\max}<0.75$, the peak value of the cross-correlation
coefficient $r(\tau)$ as defined by \citealt{edelson2019first})
reported by \citet{edelson2019first}, confirming that disc reprocessing, not coronal X-ray heating, governs the UV/optical lag structure recovered here.

\subsection{Comparison with JAVELIN and ICCF}

Although the primary goal of this work is to assess the ability of the KSG mutual information estimator to recover inter-band reverberation lags, it is also instructive to compare its performance against the two most widely adopted techniques in AGN reverberation mapping, namely the Interpolated Cross-Correlation Function (ICCF) and JAVELIN.

For each UV/optical band pair, we computed the absolute discrepancy between the MI lag estimate and the JAVELIN lag and the corresponding discrepancy between ICCF and JAVELIN, Eq.~\ref{D1} and Eq.~\ref{D2}, respectively.
\begin{equation}
D_{\rm MI} = |\tau_{\rm MI} - \tau_{\rm JAV}|,
\label{D1}
\end{equation}
\begin{equation}
D_{\rm ICCF} = |\tau_{\rm ICCF} - \tau_{\rm JAV}|.
\label{D2}
\end{equation}

The mean absolute discrepancy was calculated, as shown in Eq.~\ref{D3}
\begin{equation}
\langle D \rangle =
\frac{1}{N}
\sum_{i=1}^{N} D_i ,
\label{D3}
\end{equation}
where $N$ is the number of UV/optical lag measurements. The median discrepancy was obtained from the ordered set of $D_i$ values, while the root-mean-square (RMS) discrepancy was computed as shown in Eq.~\ref{D4}
\begin{equation}
D_{\rm_{RMS}}=\sqrt{\frac{1}{N}\sum_{i=1}^{N}D_i^2}.
\label{D4}
\end{equation}

The resulting statistics are presented in Table~\ref{comparison_javelin}.

\begin{table}
\centering
\caption{Comparison between MI and ICCF relative to the JAVELIN lag estimates for the 20 UV/optical band pairs.}
\label{comparison_javelin}
\begin{tabular}{lccc}
\hline
Method & Mean (d) & Median (d) & RMS (d) \\
\hline
MI - JAVELIN   & 0.151 & 0.084 & 0.267 \\
ICCF - JAVELIN & 0.180 & 0.143 & 0.240 \\
\hline
\end{tabular}
\end{table}

The MI estimator exhibits smaller mean and median discrepancies relative to JAVELIN than the ICCF. The mean absolute difference decreases from 0.180~d for ICCF to 0.151~d for MI, corresponding to a reduction of approximately 16\%. Likewise, the median discrepancy decreases from 0.143~d to 0.084~d, corresponding to a reduction of approximately 41\%, indicating that the typical MI lag estimate lies substantially closer to the JAVELIN solution.

A complementary comparison was performed on a case-by-case basis. Among the 20 UV/optical lag measurements analysed in this work, the MI estimator produced lag values closer to the JAVELIN estimates in 14 cases (70\%), while ICCF was closer in only 6 cases (30\%). This tendency suggests that the MI estimator generally reproduces the JAVELIN solutions more faithfully than ICCF.

To assess whether this apparent advantage is statistically significant, we applied a paired Wilcoxon signed-rank test to the two discrepancy distributions. The resulting $p$-value ($p=0.165$) does not allow rejection of the null hypothesis that both methods perform equivalently at the conventional 95\% confidence level. Nevertheless, the systematic reduction in both the mean and median discrepancies, together with the 70\% agreement rate, indicates a consistent trend favouring MI.

This result is particularly relevant because JAVELIN is generally regarded as one of the most sophisticated lag-recovery methods currently available, relying on Bayesian inference and stochastic process modelling of AGN variability. While JAVELIN often provides highly robust lag estimates, its application can become computationally expensive when analysing large samples, densely sampled light curves, or extensive Monte Carlo experiments.

In contrast, the KSG mutual information estimator is comparatively simple to implement, computationally efficient, and free from assumptions regarding linear correlations between the underlying signals. The present results therefore suggest that MI may represent a practical alternative to ICCF when a computationally inexpensive estimator is desired, while still maintaining close agreement with the lag solutions obtained by JAVELIN.

The closer agreement between MI and JAVELIN is physically consistent with both being less sensitive to the extended, diffuse-continuum-driven tail of the transfer function that biases the ICCF centroid \citep{cackett2018accretion}. This suggests that peak/model-based lag estimators recover a delay closer to the direct disc-reprocessing signal, while the ICCF centroid is more strongly weighted by BLR diffuse continuum contamination, particularly in high-luminosity sources such as Mrk~509.

Although the current sample is too small to establish a statistically significant superiority of MI over ICCF, the observed tendency motivates future investigations using larger AGN samples. If confirmed, this behaviour would support the use of mutual-information-based lag recovery as a competitive and computationally efficient tool for reverberation mapping studies.

\subsection{Comparison with the multifractal characterisation}
\label{subsec:mfdma_discussion}
 
\citet{Souza2026RM} characterised the same eight light curves of
these four sources using Multifractal Detrended Moving Average (MFDMA), reporting for each band the
multifractal spectral width $\Delta\alpha$, the asymmetry parameter
$A$, the Hurst exponent $H$, and the fractional variability,
$F_{\rm var}$, together with a superlinear scaling
$\Delta\alpha \propto F_{\rm var}^{1.69}$ across the full sample.
Their band-by-band picture is markedly source-dependent. In
Mrk~509, every band except B and V is multifractal, most strongly
HX and SX ($\Delta\alpha = 0.216$ and $0.321$); in NGC~5548,
multifractality is again strongest in HX and SX, with SX reaching
the highest $\Delta\alpha = 0.681$ in the whole sample and the only
near-antipersistent Hurst exponent ($H = 0.015$), while B and V are
near-monofractal. NGC~4593 follows a similar pattern, HX and SX
more multifractal than the optical bands, with W2 and M2 strongly
right-skewed ($A = 2.456$ and $2.836$) and B, V essentially
monofractal. NGC~4151 stands apart: only HX is clearly multifractal
($\Delta\alpha = 0.542$, left-skewed with $A = 0.264$), while every
other band, including SX, is near-monofractal
($\Delta\alpha \leq 0.04$).
 
Comparing this with the MI-ICCF discrepancies in
Table~\ref{tab:kernelmi_lags}, the clearest qualitative link is in
the UV/optical regime: among the five subsidiary bands compared to
W2 in this work (M2, W1, U, B, V), M2 and W1 carry the highest and
second-highest $\Delta\alpha$ in three of the four sources
\citep{Souza2026RM}, and these are also the bands with the smallest
mean MI-ICCF offsets ($\langle|\Delta\tau|\rangle \simeq 0.13$ and
$0.06$\,d; Section~\ref{subsec:byband}), consistent with their
comparatively structured, higher-amplitude variability producing a
better-defined MI($\tau$) peak. The X-ray bands break
this correspondence: NGC~5548/SX and NGC~4151/HX have the two
highest $\Delta\alpha$ values in the entire sample
\citep{Souza2026RM}, yet their MI-ICCF discrepancies are among the
largest found here ($|\Delta\tau| = 1.90$ and $3.83$\,d). The
exception is NGC~4593, whose HX and SX bands are simultaneously
among the most multifractal in the sample \citep{Souza2026RM} and
the only X-ray bands for which the MI lag agrees with the ICCF to
within $0.16$\,d (Section~\ref{subsec:res_4593}). Taken together,
this suggests that $\Delta\alpha$ alone does not determine MI
performance: in the UV/optical regime higher multifractality
appears to coincide with more reliable lag recovery, but in the
X-ray bands the outcome instead seems to depend on source-specific
properties, such as the X-ray/UV coupling strength
\citep{edelson2019first}, that a single complexity measure does not capture. A direct, band-by-band statistical test of this relation across the full 28-entry table is left for future work.

The UV/optical correspondence between higher $\Delta\alpha$ and better MI recovery suggests that stronger, more structured disc-driven variability yields a sharper, more identifiable MI peak. The breakdown of this relation in the X-ray bands indicates that multifractality alone does not encode reprocessing efficiency; instead, source-specific X-ray/UV coupling strength \citep{edelson2019first} remains the dominant factor, implying $\Delta\alpha$ is a UV/optical-specific, not universal, diagnostic for MI-based lag recovery.

% ==================================================================
\section{Conclusions}
\label{sec:conclusion}
% ==================================================================

We have applied KSG MI lag estimation ($k=5$,
150-iteration RSS bootstrap) to the multiwavelength light curves of
four Seyfert AGNs from the \textit{Swift} IDRM survey of
\citet{edelson2019first}, providing a systematic comparison of
MI-based reverberation mapping against both the ICCF and JAVELIN lags
on a multi-source, multi-band dataset. In the UV/optical regime the
method recovers inter-band lags consistent with the ICCF and JAVELIN
across all four sources, with the closest agreements in the M2 and W1
bands ($|\Delta\tau| \leq 0.17$\,d on average) and degrading
performance toward the optical B and V bands, where low variability
amplitude reduces the discriminating power of the MI peak. The best
single-band agreement in the full sample is NGC~5548/B
($|\Delta\tau| = 0.03$\,d), and the MI recovers the correct lag sign
for all UV/optical bands in Mrk~509. These results provide an
observational validation, on real multi-band AGN data, of the
proof-of-concept put forward by \citet{Chakraborty2022} using toy
models, and extend the scope of alternative randomness-based
estimators explored by \citet{Chelouche2017} to the multiband AGN
disc-reprocessing regime.

A complementary comparison with JAVELIN further supports the potential of the MI approach. Across the 20 UV/optical band pairs analysed, MI lag estimates were closer to the JAVELIN solutions in 14 cases (70\%), compared to 6 cases for ICCF. Moreover, the mean and median MI-JAVELIN discrepancies (0.151\,d and 0.084\,d, respectively) are smaller than the corresponding ICCF-JAVELIN values (0.180\,d and 0.143\,d). Although the current sample does not provide statistically significant evidence for the superiority of MI over ICCF, these results indicate that MI is capable of reproducing JAVELIN-like lag estimates while retaining a comparatively simple and computationally efficient implementation.

In the X-ray bands the MI lags deviate substantially from the ICCF
reference for three of the four sources, in several cases with
opposite sign. The discrepancies reach $|\Delta\tau| = 5.86$\,d for
Mrk~509/HX, $3.83$\,d for NGC~4151/HX, $3.77$\,d for NGC~5548/HX, and
$1.90$\,d for NGC~5548/SX. The exception is NGC~4593, for which both
X-ray lags agree with the ICCF within the bootstrap uncertainty
($|\Delta\tau| \leq 0.13$\,d). These failures are consistent with the
weak X-ray/UV physical coupling reported by \citet{edelson2019first},
who find $r_{\rm max} < 0.75$ for all eight X-ray/UV pairs versus
$r_{\rm max} > 0.8$ for all twenty UV/optical pairs, indicating that
the X-ray variability is not straightforwardly connected to the UV
through the disc-reprocessing channel that governs the UV/optical
lags.

The multifractal characterisation of the same light curves by
\citet{Souza2026RM} provides a band-resolved framework for
understanding these outcomes. The MI-ICCF agreement is best in bands
with moderate to high multifractal spectral width $\Delta\alpha$ and
degrades as $\Delta\alpha$ approaches zero in near-monofractal optical
bands with low $F_{\rm var}$. In the X-ray bands, high $\Delta\alpha$,
reaching $0.681$ for NGC~5548/SX, the largest in the sample, which
also carries the only near-antipersistent Hurst exponent ($H = 0.015$)
reported by \citet{Souza2026RM}, does not on its own improve MI
performance, with NGC~4593 as the notable exception. Taken together,
these results suggest $\Delta\alpha$ as a candidate diagnostic for
anticipating where MI-based lag estimation is likely to succeed in AGN
multiband datasets, particularly in the UV/optical regime, while
source-specific factors remain important for the X-ray bands.

Physically, these results reinforce the standard disc-reprocessing picture \citep{shakura1973black, cackett2007testing}: MI reliably recovers UV/optical lags where the disc continuum dominates, while previous work has shown that lags in some bands, most notably the U band, are longer than predicted by the thin-disc model, implying either a larger disc size than expected or an additional contribution from the BLR and diffuse continuum emission \citep{cackett2018accretion, edelson2019first}. The pervasive X-ray failure confirms the weak physical X-ray/UV coupling reported by \citet{edelson2019first}, indicating that coronal X-ray variability is not simply reprocessed into the UV/optical bands through the same channel governing the disc response.

The main current limitation of the MI-based approach presented here is not the size of the sample but its homogeneity. Because $\widehat{\mathrm{MI}}(\tau;k)$ requires both light curves to  be evaluated on a common time grid, irregular sampling must be bridged by interpolation, and the fraction of interpolated points needs to remain minimal to avoid introducing spurious structure into the $\mathrm{MI}(\tau)$ curve. The sample analysed here satisfies this requirement reasonably well, which allowed a first successful  application of the method to reverberation-mapping data. Applying this technique to other AGN samples, spanning a wider range of sampling cadences and data quality, will allow us to systematically characterise the regime in which the MI estimator remains reliable.

\section*{Acknowledgements}
% ==================================================================
RS acknowledges the Universidade de Pernambuco (UPE) for its support
and resources. This work was carried out within the activities of the Grupo de Ensino das Ci\^encias e Matem\'aticas, UPE. We also acknowledge the Graduate Program in Applied Physics (PPGFA) at the Federal Rural University of Pernambuco (UFRPE) for its support and resources throughout this research. DOF acknowledges the Universidade Federal do Rio Grande do Norte (UFRN) for its support and resources, and CNPq for financial support. LBTS acknowledges the Group of Orbital \& Rotational Research in Irregular Objects \& Observational astroNomy (ORION), UNESP, and the PostGrad Program in Systems Engineering (PPGES), Universidade de Pernambuco (UPE), for their support and resources. MVDS acknowledges the Departamento de Qu\'imica Fundamental, Universidade Federal de Pernambuco (UFPE), for its support and resources. TS acknowledges the Departamento de Estat\'istica e Inform\'atica, Universidade Federal Rural de Pernambuco (UFRPE), for its support and resources. EH has received funding from the European Union's Horizon Europe research and innovation programme under grant agreement No.\ 101188037 (AtLAST2). The IAC is an accredited Severo Ochoa Centre of Excellence supported by grant CEX2025-001609-S, funded by MICIU/AEI/10.13039/501100011033. This research has made use of the VizieR catalogue access tool, CDS, Strasbourg, France (DOI: 10.26093/cds/vizier). The original description of the VizieR service was published in \citet{ochsenbein2000}, A\&AS, 143, 23.

\section*{Data Availability}

The data used in this study are available in the VizieR catalogue at
\url{https://doi.org/10.26093/cds/vizier.18700123}. The mutual
information estimator used in this work is the
\texttt{mutual\_info\_regression} implementation of \textsc{scikit-learn},
documented at
\url{https://scikit-learn.org/stable/modules/generated/sklearn.feature_selection.mutual_info_regression.html}.

\bibliographystyle{mnras}
\bibliography{example}

\appendix
\onecolumn

\section{Time Series for each AGN}

{\centering
\noindent\includegraphics[width=\textwidth,height=0.80\textheight,keepaspectratio]{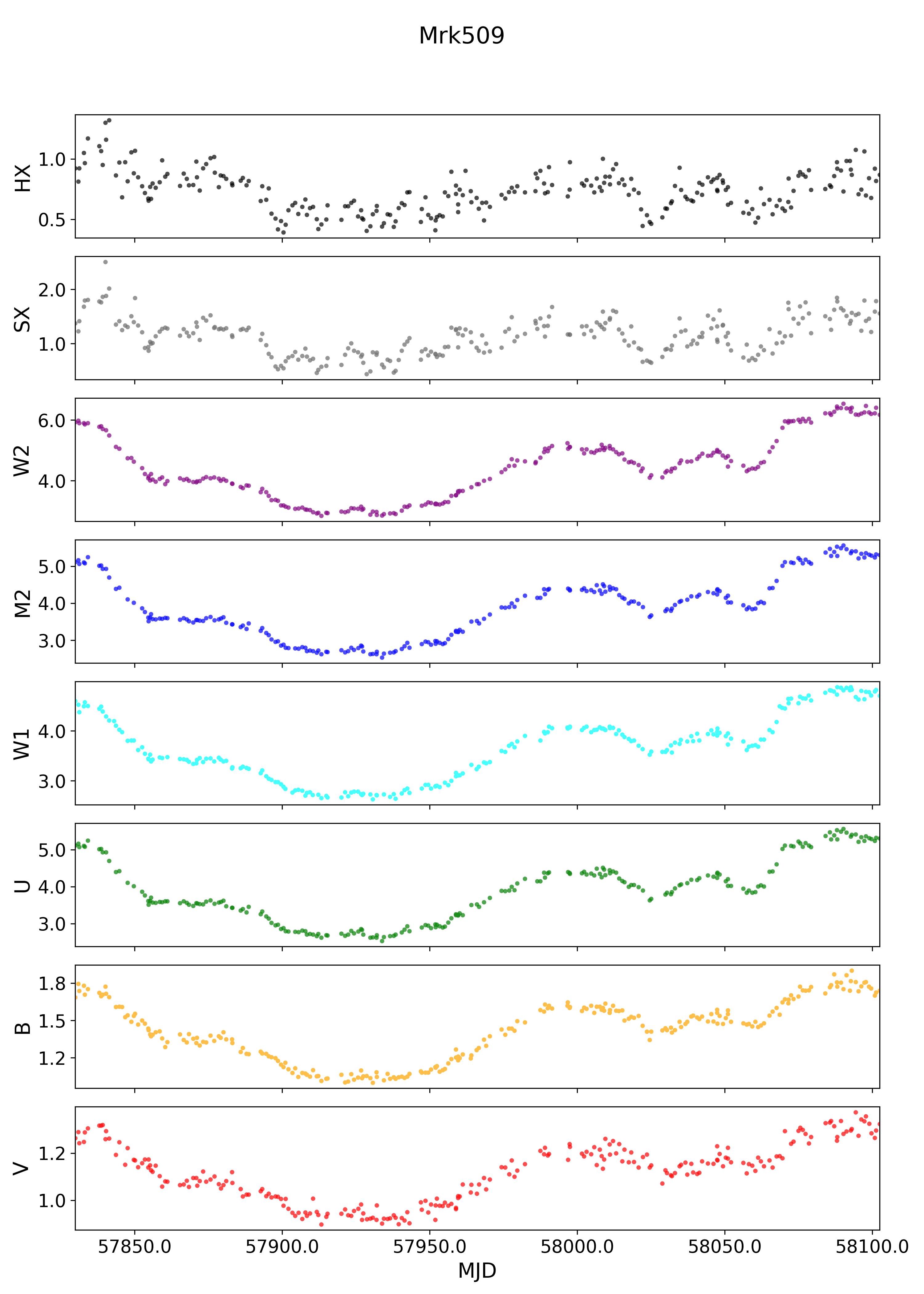}
\captionof{figure}{The original light curve of Mrk~509 for all eight
observational bands, ordered by wavelength from top to bottom. The top
two panels correspond to XRT observations in the HX (1.5,10 keV) and
SX (0.3,1.5 keV) bands, while the lower six panels show the UVOT
bands (W2, M2, W1, U, B, and V). Each panel displays the flux as a
function of MJD, sharing a common time axis but with independent flux
scales. X-ray fluxes are given in units of counts s$^{-1}$, and UVOT
fluxes are expressed in units of
$10^{-14}$\,erg\,cm$^{-2}$\,s$^{-1}$\,\AA$^{-1}$.}
\label{fig:lc_plot_Mrk509}

\noindent\includegraphics[width=\textwidth,height=0.80\textheight,keepaspectratio]{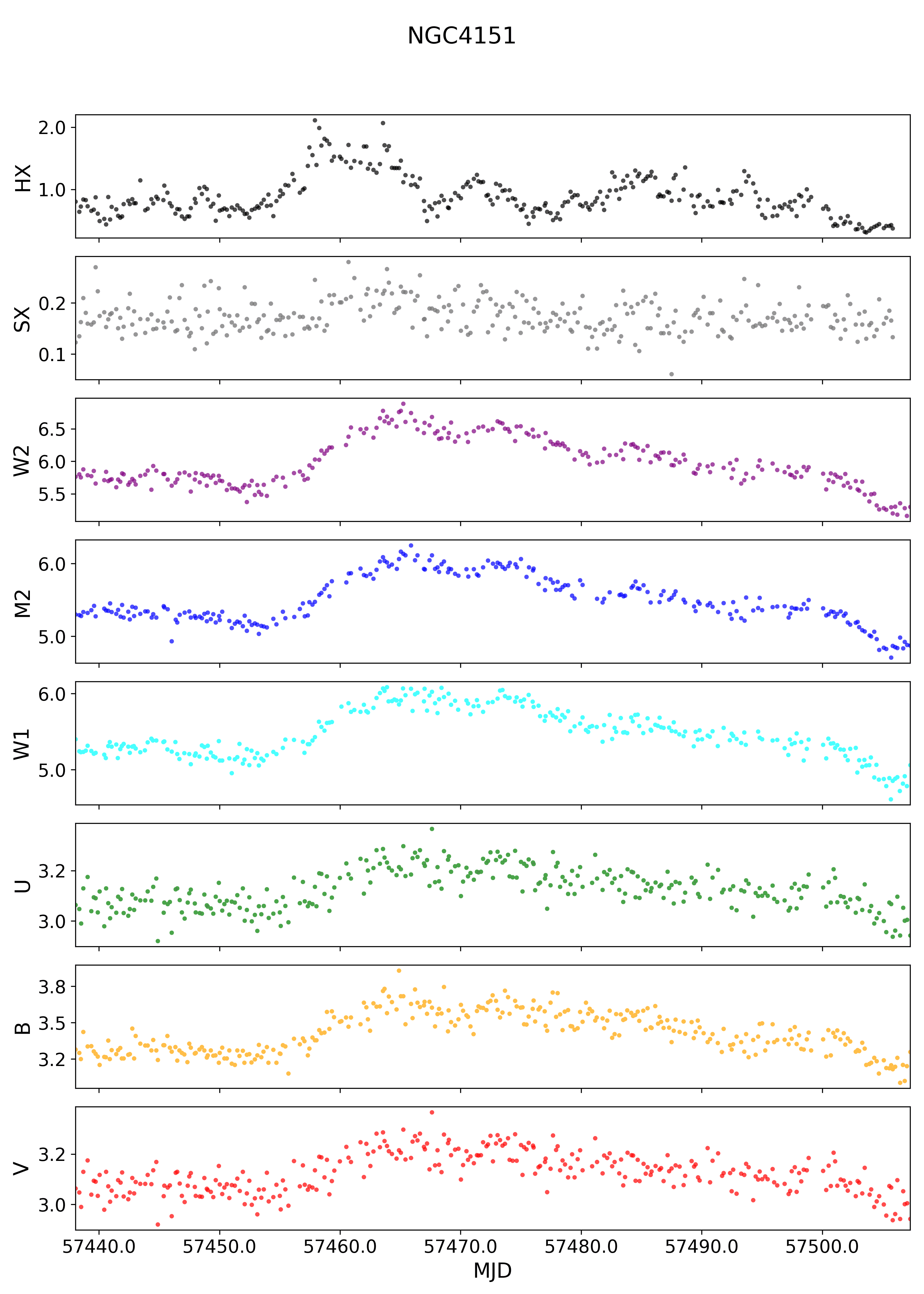}
\captionof{figure}{Same as Fig.~\ref{fig:lc_plot_Mrk509} for NGC~4151.}
\label{fig:lc_plot_NGC4151}

\noindent\includegraphics[width=\textwidth,height=0.80\textheight,keepaspectratio]{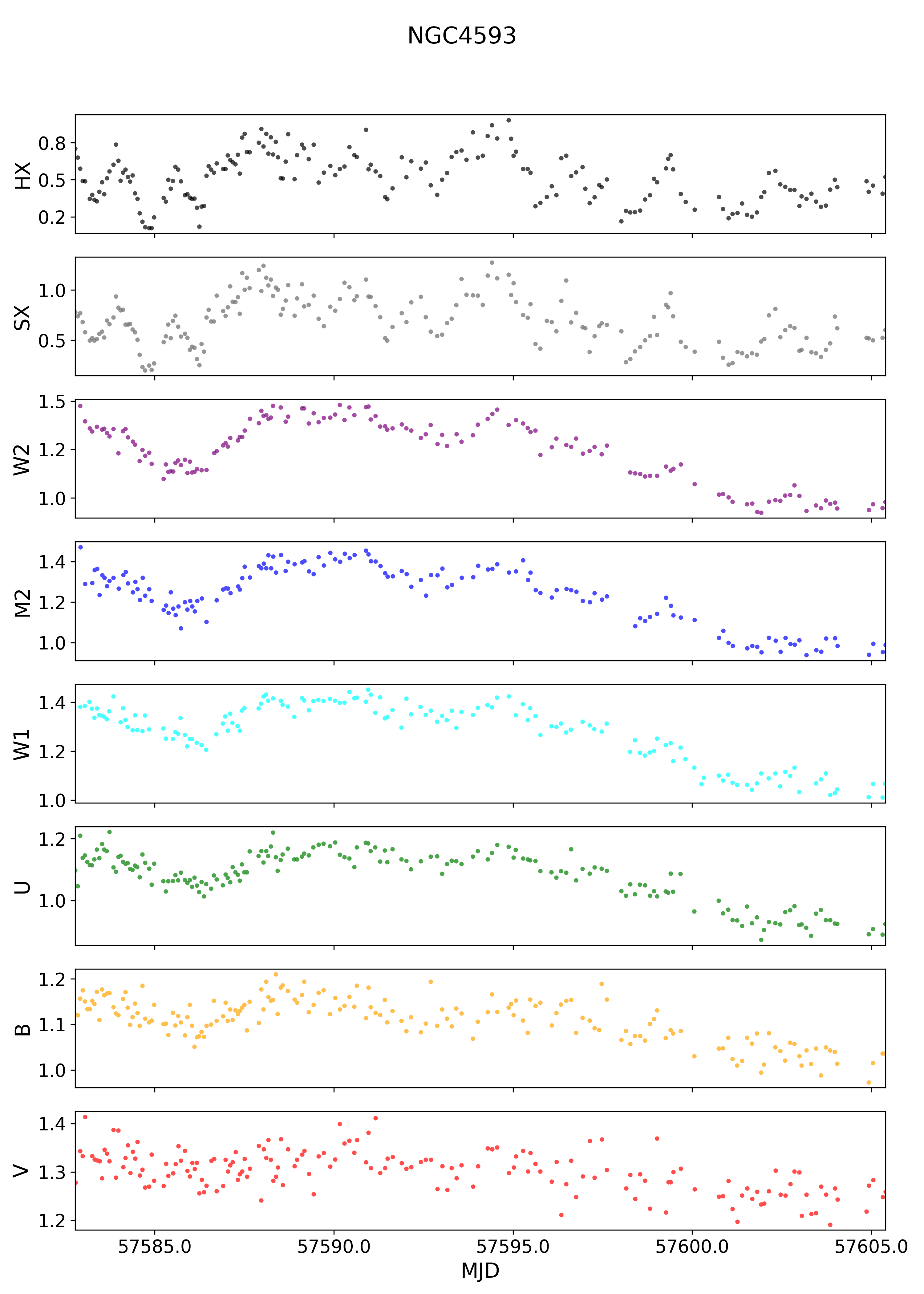}
\captionof{figure}{Same as Fig.~\ref{fig:lc_plot_Mrk509} for NGC~4593.}
\label{fig:lc_plot_NGC4593}

\noindent\includegraphics[width=\textwidth,height=0.80\textheight,keepaspectratio]{figures/AGN_Light_Curves/NGC5548_scatter_lightcurves.png}
\captionof{figure}{Same as Fig.~\ref{fig:lc_plot_Mrk509} for NGC~5548.}
\label{fig:lc_plot_5548}
\par}

\end{document}